\documentclass[sigconf]{acmart}

\usepackage{booktabs} % For formal tables
\usepackage{ragged2e}
\usepackage{xspace}
\usepackage{graphicx}
\usepackage{svg}
\usepackage{amsmath}
\usepackage{adjustbox}
\usepackage{color}
\usepackage{refstyle}
\usepackage{caption}
\usepackage{multirow}

\renewcommand\footnotetextcopyrightpermission[1]{} % removes footnote with conference information in first column

\usepackage{balance}

\begin{document}
\nocite{*}
% remove comment  to show references that were not cited 
%%%%% Commands
\renewcommand{\texttt}[1]{%
  \begingroup
  \ttfamily
  \begingroup\lccode`~=`/\lowercase{\endgroup\def~}{/\discretionary{}{}{}}%
  \begingroup\lccode`~=`[\lowercase{\endgroup\def~}{[\discretionary{}{}{}}%
  \begingroup\lccode`~=`.\lowercase{\endgroup\def~}{.\discretionary{}{}{}}%
  \catcode`/=\active\catcode`[=\active\catcode`.=\active
  \scantokens{#1\noexpand}%
  \endgroup
}
  
\newcommand{\one}{({\em i}\/)}
\newcommand{\two}{({\em ii}\/)}
\newcommand{\three}{({\em iii}\/)}
\newcommand{\four}{({\em iv}\/)}
\newcommand{\five}{({\em v}\/)}
\newcommand{\six}{({\em vi}\/)}
\newcommand{\seven}{({\em vi}\/)}

\def\tofill{\textcolor{red}{XX}\xspace}
\def\tofillp{\textcolor{red}{XX\%}\xspace}
\def\tofilltext{\textcolor{orange}{FILL-TEXT}\xspace}

\def\eg{\emph{e.g., }\xspace}
\def\etc{\emph{etc.}\xspace}
\def\ie{\emph{i.e.,}\xspace}
\def\etal{\emph{et al.}\xspace}
\def\vs{\emph{vs.}\xspace}
\def\cf{\emph{cf.}\xspace}
\definecolor{wscolor}{RGB}{26, 26, 255}
\newcommand\ws[1]{  \textcolor{wscolor}{#1}} %for websites
%for the targeted vs blanket conundrum
\def\troll{dispersed \xspace}
\def\trollnot{focused \xspace}

%\makeatletter
%\newcommand{\myconst}[1]{#1\renewcommand{\@currentlabel}{#1}}
%%%%%%%% Data by Numbers Dataset %%%%%%%%%
\def\TotalNumberOfNotices{1,885,267}
\def\TotalNumberOfInfringingURLs{1,054,248,823}
\def\TotalNumberOfNoticeSenders{38,523}
\def\TotalNumberOfNoticeRecipient{223} %% includes notice sender null
\def\TotalNumberOfNoticePrincipal{20,686}
\def\TotalNumberOfNoticeTypes{9}

\def\TotalNumberOfDomains{1,090,173}
\def\TotalNumberOfIPAddress{206,863}
\def\TotalNumberOfCountries{165}
\def\TotalNumberOfContinents{8}
\def\TotalNumberOfAS{4,049}
\def\TotalNumberOfServers{137,688}

\def\DistinctNumberOfNoticeTypes{8}

\def\NumberOfTwitterURL{252,051}
\def \NumberOfTwitterHandles{148,572}

\def\NumberOfURLsWeeklyCheck{2,094,847}
\def\TransitionTwoHunderedtoTimeout{13,938}
\def\TransitionTwoHunderedtoFourHundredtoTimeout{406}

\def\commentsoff{1}

\ifx\commentsoff\undefined
	\newcommand\gareth[1]{\textbf{\textcolor{blue}{GAR: #1}}}
	\newcommand\dami[1]{\textbf{\textcolor{red}{DAM: #1}}}
    \newcommand\ignacio[1]{\textbf{\textcolor{orange}{IGN: #1}}}
    \newcommand\gianluca[1]{\textbf{\textcolor{violet}{GIAN: #1}}}
     \newcommand\steve[1]{\textbf{\textcolor{purple}{STEVE: #1}}}
     \newcommand\ben[1]{\textbf{\textcolor{green}{BEN: #1}}}
\else
	\newcommand\gareth[1]{}
	\newcommand\dami[1]{}
    \newcommand\ignacio[1]{}
    \newcommand\gianluca[1]{}
    \newcommand\steve[1]{}
    \newcommand\ben[1]{}
\fi

\title{Who Watches the Watchmen: \\ Exploring Complaints on the Web}

%\title{Not \emph{You} Again! \\ Exploring Complaints on the Web}

\author{Damilola Ibosiola}
\affiliation{%
  \institution{Queen Mary University of London}
}
\email{d.i.ibosiola@qmul.ac.uk}

\author{Ignacio Castro}
\affiliation{%
  \institution{Queen Mary University of London}
}
\email{i.castro@qmul.ac.uk}

\author{Gianluca Stringhini}
\affiliation{%
  \institution{Boston University}
}
\email{gian@bu.edu}

\author{Steve Uhlig}
\affiliation{%
  \institution{Queen Mary University of London}
}
\email{steve.uhlig@qmul.ac.uk}

\author{Gareth Tyson}
\affiliation{%
  \institution{Queen Mary University of London}
}
\email{g.tyson@qmul.ac.uk}

\renewcommand{\shortauthors}{D. Ibosiola et al.}

\begin{abstract}

% The global nature of the web crosses jurisdictions making its regulation notoriously difficult. 
Under increasing scrutiny, many web companies now offer bespoke mechanisms allowing any third party to file complaints (\eg requesting the de-listing of a URL from a search engine).
While this self-regulation might be a valuable web governance tool, it places huge responsibility within the hands of these organisations that demands close examination.
We present the first large-scale study of web complaints (over 1 billion URLs). We find a range of complainants, largely focused on copyright enforcement. Whereas the majority of organisations are occasional users of the complaint system, we find a number of bulk senders specialised in targeting specific types of domain. We identify a series of trends and patterns amongst both the domains and complainants. By inspecting the availability of the domains, we also observe that a sizeable portion go offline shortly after complaints are generated. This paper sheds critical light on \emph{how} complaints are issued, \emph{who} they pertain to and \emph{which} domains go offline after complaints are issued.

\end{abstract}

\maketitle

%%%%%%%%%%%%%%%%%%%%%%%%%%% ACTUAL TEXT: Begin %%%%%%%%%%%%%%%%%%%%%%%%%%%%

%%%%%%%%%%%%%%%%%%%%%%%%%%%%%%%%%%%%%%%%%%%%%%%%%%%%%%%%%%%%%%%%%
\section{Introduction}
%%%%%%%%%%%%%%%%%%%%%%%%%%%%%%%%%%%%%%%%%%%%%%%%%%%%%%%%%%%%%%%%%

The web has proven a powerful platform for the large-scale distribution of content. 
Notoriously difficult to regulate, individual web organisations have been frequently left to decide how to best handle issues related to legal, regulatory and even moral matters, \eg moderation of online discourse, removal of copyright infringing content, mitigation of online harassment.
This has generated significant societal attention, with major companies like Google and Twitter coming under increasing public scrutiny~\cite{giannoumis2014regulating,stiegler2013regulating}. 
Consequently, large web organisations have begun to implement bespoke mechanisms to allow third parties to register complaints. For example, Google's complaint system allows anybody to issue notices requesting the removal of specific results from their search listings, whereas Twitter enables users to report posts they believe to be infringing policy (\eg because of bot activity~\cite{gilani2017bots}).
Although a valuable tool in the wider landscape of web governance, this places considerable responsibility within the hands of these organisations, who must decide which complaints are legitimate and how they should be dealt with (often referred to as self regulation~\cite{urist2006s}). Yet, to date, we have little evidence regarding how these complaint procedures are handled, how successful they are, or who they are targeted against and by whom. 

We argue that the scale of these complaints and their impact on the wider society's perception of the web, warrant detailed investigation. 
There are multiple interesting questions, such as who generates complaints? Whom do these complaints pertain to? What are the characteristics of  domains that receive complaints? Does content remain online after complaints? 
With these questions in mind, this paper presents a \emph{large-scale study of web complaints}. 
Building on past work within the legal domain~\cite{Urban2016,Urban2005,Seng2014}, we have gathered complaints from hundreds of transparency reports made available by organisations including Google, Vimeo, Bing and Twitter, and reflected in the Lumen database (\S\ref{sec:Dataset}). These reports expose detailed information about web complaints  covering over 1 billion URLs ranging from copyright infringement to governmental notices. 

We start our analysis by characterising the nature and scale of organisations that generate notices (complainants). 
Despite the presence of numerous complaint categories, copyright notices largely dominate, with 98.6\% of all complaints (\S\ref{sub:what_is_injected}).
A critical minority of organisations play a remarkably prominent role in this, with the top 10 alone contributing 41\% of all notices sent (\S\ref{sub:who_injects}). 
Our analysis reveals that the majority of \emph{notices} are generated by large content-based organisations (\eg NBC, Fox). Despite this, we find that the majority of notice senders are occasional users of the complaint system: 94\% of complainants issue fewer than 100 notices.
These different groups tend to rely on different types of complaints. 
For example, large copyright enforcers (\eg Rivendell) generate millions of copyright notices, whereas governmental agencies (\eg Roskomnadzor) issue a far smaller set of targeted court and governmental notices. This leads us to focus on the categories of websites that these different types of notices pertain to (\S\ref{sub:topics}). 
We find that notices are heavily biased towards certain types of website. For example, websites categorised as `File Sharing' and `Elevated Exposure' are hugely over-represented amongst our complaint dataset. Similarly, individual notice senders tend to target specific topics, \eg 60\% of complaints by \textit{Cam Model Protection} target adult websites, whereas 55\% by \textit{NBCUniversal} target Entertainment, shedding light on the priorities of organisations utilising the complaint services. 

This leads us to investigate if equivalent dynamism exists on the part of the reported websites themselves (\S\ref{sec:characterising_reported_domains}). We find that a small (and constantly evolving) set
of websites dominate the ranking of the most reported, with a few domains that remain prominent throughout our entire measurement period: the top 1\% of domains accumulate 63\% of complaints (\S\ref{sub:hot_domains}). These include many obscure websites, which are unlikely to be widely known, \eg 22 out of the top 30 most frequently reported websites are not even in the Alexa Top 100K, and the correlation between the popularity of domains in terms of notices and Alexa rank is just 0.13.
Deeper analysis reveals that these trends are dictated by the activity of a small set of \emph{extremely} aggressive notice senders (\S\ref{sub:domain_stability}), whose `bursty' behaviour creates high levels of instability.
% Complainers frequently produce notices in outbursts  and in the most extreme cases, the statistical variance exceeds 100K complaints per day. 
% For instance, 6 out of the top 10 most reported domains receive at least 98\% of their complaints from a single organisation. 
% This generalises across the wider dataset, and creates huge instability in the day-to-day ranking of top reported domains. 

We finally inspect the availability of the webpages that are complained about, with the conjecture that websites receiving many complaints may be more likely to go offline (\S\ref{sub:target_behaviour}). We confirm that many are highly ephemeral: 22\% of all domain names soon get taken offline (NXDOMAIN), whereas our HTTP liveness checks show that a further 19\% of all resources fail to return 200 OK responses within 1 week of a complaint. We correlate these with a number of factors to find that the `success' rate of complainants differs dramatically, with the most successful (Rivendell) seeing 55\% of its complaints acted upon in contrast to others complainants where the figure is below 1\%.
The published version of this report is available at~\cite{ibosiola2019watches}.

%We also observe similar trends across website categories, \eg 24\% of Elevated Exposure web servers cease accepting TCP connections compared to just 9\% for Blogs. 
%Finally, we summarise our key findings and future work (\S\ref{sec:conclusion})

%%%%%%%%%%%%%%%%%%%%%%%%%%%%%%%%%%%%%%%%%%%%%%%%%%%%%%%%%%%%%%%%%
\section{Methodology \& Dataset}\label{sec:Dataset}
%%%%%%%%%%%%%%%%%%%%%%%%%%%%%%%%%%%%%%%%%%%%%%%%%%%%%%%%%%%%%%%%%

This section presents our data collection methodology. This has two goals: \one~To provide vantage into a broad set of web complaints, covering enough domains to provide meaningful insight; and \two~To annotate these complaints with sufficient metadata to shed light on domain activities and topics.

%%%%%%%%%%%%%%%%%%%%%%%%%%%%%%%%%%%%%%%
\subsection{Website Complaints}
%%%%%%%%%%%%%%%%%%%%%%%%%%%%%%%%%%%%%%%

First, we detail our methodology for gathering complaints issued about websites. Naturally, it is impossible to get complaints issued to \emph{all} websites, because the vast majority do not make this information available. Hence, we focus on sites that expose transparency reports, \eg Bing, Twitter, Google, Vimeo. These reports list complaints received by the organisations, including relevant metadata. To gather these, we utilise Lumen\footnote{\url{https://lumendatabase.org/}} --- a database run by the Berkman Klein Center for Internet \& Society. It aggregates and publishes transparency report data pertaining to complaints issued towards 223 organisations. The exact purpose of each complaint differs, \eg a complaint to Bing will normally request the removal of search results, whilst complaints to Vimeo will concern the removal of videos. However, each notice (\ie complaint) includes the following standard fields:

\begin{itemize}
\item \emph{Notice Type:} the category of notice which has been reported. For example, a Digital Millennium Copyright Act (DMCA), trademark or data protection \etc 

\item \emph{Notice Sender:} or complainant, is the organisation who submitted the notice. %Throughout the rest of the paper we also refer to notice senders as complainants.

\item \emph{Notice Recipient:} the web publisher or service provider where the infringing notice is sent to. 

\item \emph{Notice Principal:} in cases of a copyright-related notice, this is the person or organisation that holds the copyright on the content reported. 

\item \emph{Infringing URL(s):} the list of URL(s) that the notice sender is requesting to be dealt with. Note that this is not necessarily a set of URLs owned by the recipient, \eg Bing may receive complaints requesting the removal of a third-party URL from its search results. 

\end{itemize}

%% Percentage of notices issued to recipients

\noindent We collected all complaints from the Lumen database between 01/01/2017 and 31/12/2017 using their API. 
In total, we extract \TotalNumberOfInfringingURLs~URL complaints from \TotalNumberOfNoticeSenders~ notice senders. 
% Complaints mostly pertain to third party URLs which the sender is asking the recipient to de-list from their site. For context, 95\% of notices are sent to search engines (primarily Google and Bing), and 4.6\% are sent to social networks (primarily Twitter and Periscope). The remainder are a mix of online organisations specialising in File Hosting (0.4\%), Education (< 0.01\%), News (< 0.01\%), and Blogs (< 0.01\%).
%we repeat this in the next section

%and Private Individual --- 0.015\%.
%We also have 197 notices with no recipients, so I could not classify them}

%These complaints cover XXXX unique URLs, hosted by \TotalNumberOfDomains~distinct domains.  

%to \TotalNumberOfNoticeRecipient~notice recipient. Google is the most frequently regarding \TotalNumberOfNoticePrincipal~copyright owners.

%%%%%%%%%%%%%%%%%%%%%%%%%%%%%%%%%%%%%%%%%%%%%%
\subsection{Website Metadata}
%%%%%%%%%%%%%%%%%%%%%%%%%%%%%%%%%%%%%%%%%%%
Once we extract the reported URLs, we compile further metadata. This section explains the metadata collected.

\vspace{4pt}
\noindent\textbf{Website Categorisation.} We classify each domain using the VirusTotal API.\footnote{\url{https://www.virustotal.com/}}
This API has been used in a wide set of research, and is known to provide good accuracy~\cite{kim2015detecting,wang2017droid,ikram2019chain}. 
The API provides a classification for each domain in our dataset, \eg games, education, file sharing, blogs \etc We later use this to understand the types of complaints generated. 
Due to the usage limitations of the API, we only categorise the top 240K domains with the most reported URLs. Note that 22\% of these could not be categorised by VirusTotal. We further annotate each domain with its Alexa ranking to gain insight into its global popularity.

\vspace{4pt}
\noindent\textbf{DNS Probes.} We utilise the Domain Name Service (DNS) to map the domains to their respective DNS records on 29/07/2018. We performed queries (IPv4), which yield 849,023 responses and 206,863 IP addresses. 
We use this data to check if the domain name is still live.

\vspace{4pt}
\noindent\textbf{Webpage Probes.} For each URL, we download its HTML and parse it to extract all embedded domains. This allows us to identify mirrors of websites hosted on multiple domains/URLs, by comparing the HTML contents. Tests are performed from a university campus, where we have confirmed no web filtering is performed. 
In total, we collect the HTML structure for 770,737 webpages. \gareth{Is this because we don't have the full 1M scrapes yet, or something different?}

\vspace{4pt}
\noindent\textbf{Liveness Checks.} We also perform periodic checks on the domains and URLs to verify if the websites are still active (\ie returning an HTTP 200 status code). 
This allows us to explore the potential efficacy of organisations seeking to remove content. 
Due to the sheer scale of the complaints ($>$1 billion URLs), we only perform checks for 2M URLs complained about between 14/07/2018 to 17/07/2018.
Upon recording a complaint from Lumen, we added its URL to a queue and began weekly checks that executed between 18/07/2018 to 14/08/2018. We record the HTML response and HTTP status code, alongside whether or not the TCP handshake timed out. Each week, we exclude URLs that have already been deleted. 

%We first retrieve the last snapshot for each URL from the Wayback Machine.\footnote{\url{https://archive.org/web/}} The Wayback Machine archives regular snapshots from a huge number of websites. For each URL, we retrieve the last date it was recorded as live. To provide an idea of how regularly the Wayback Machine takes snapshots, Figure~\ref{fig:domain_wayback_archive_snapshot} presents the CDF of the intervals between all archived snapshots of the domains. About 80\% of domains have a snapshot taken at least once every 5 days, suggesting relatively high granularity (\ie accuracy within the bounds of approximately one week). That said, we have no guarantee that this date was the \emph{last} day the website was live, or simply the last day the website was scraped by the Wayback Machine. Hence, to complement these results we also gather finer grained data by performing regular HTTP probes. 

%%% excluded
% To streamline the crawler, after each weekly check, we exclude URLs that are already dead (\ie return 404 status) from the next check.

%\begin{figure} [t]
%\includegraphics[width=7cm]{domain_wayback_archive_snapshot}
%\caption{CDF showing intervals between dates for which a domain is archived by the Wayback Machine}
%\label{fig:domain_wayback_archive_snapshot}
%\end{figure}

\subsection{Limitations \& Ethical Considerations}

We emphasise that Lumen only provides vantage into 223 complaint recipients, obviously a small fraction of the world's online organisations. 
These consist primarily of: \emph{Google}---79\%, \emph{Bing}--15.6\%, \emph{Twitter}--3.8\%, \emph{Periscope}--0.8\% and \emph{Vimeo}--0.4\%. 
We do not assert that our findings generalise beyond these organisations, although the scale of these five companies indicates that the insights are highly valuable nevertheless.
%Second, complaints tend to contain URLs that do not belong to the complaint recipient. For example, complaints sent to Google tend to contain third party URLs (\eg \texttt{gorillavid.in}) with the requesting of de-listing it from their search results. Hence, we gain vantage into the specific URLs, but we cannot infer whether the actual owner of the URL receives the same notice. This makes is impossible to derive a causal link between a notice and a domain going offline. 
We should also highlight that our research covers websites that may participate in illegal activity (\eg copyright infringement). We restrict this collection to downloading the webpage HTML and checking server liveness, ensuring that \emph{no} measurements involve participating in a websites activities, \eg registering user accounts or downloading content. 
Note that this may still result in advertisement revenue being generated by the website. Unfortunately, this was impossible to avoid considering the nature of our measurements. That said, we limit ourselves to accessing URLs a small number of times ($<10$), minimising any potential revenue. 

%OLD PARAGRAPH BELOW
%We limit ourselves to measurements that do \emph{not} involve participating in a websites activities, \eg registering user accounts or downloading content. Instead, these measurements are restricted to downloading the webpage HTML and checking server liveness. Note that this may result in advertisement revenue being generated by the website. Unfortunately, this was impossible to avoid considering the nature of our measurements. That said, we limit ourselves to accessing URLs a small number of times ($<10$), minimising any potential revenue. 

% These are made readily available via transparency reports. Hence, we do not release or analyse data that is not already in the public domain.  

%%%%%%%%%%%%%%%%%%%%%%%%%%%%%%%%%%%%%%%%%%%%%%%%%%%%%%%%%%%%%%%%%
% \section{Characterising Complaints}
\section{complaints \& Notice Senders} 
\label{sec:characterising_complaint}
%%%%%%%%%%%%%%%%%%%%%%%%%%%%%%%%%%%%%%%%%%%%%%%%%%%%%%%%%%%%%%%%%

In this section, we investigate the senders and receivers of complaints. Specifically, we are interested in exploring \emph{who} sends complaints and \emph{what} they pertain to.

%%%%%%%%%%%%%%%%%%%%%%%%%%%%%%%%%%%%%%%%%%%
% \subsection{Characterising Notice Types}
\subsection{What Types of Complaints Exist?}
\label{sub:what_is_injected}
%%%%%%%%%%%%%%%%%%%%%%%%%%%%%%%%%%%%%%%%%%%
%% Average number of domains injected per notice = 31
%% Average number of URL injected per notice = 560
%% Number of notices to a single domain = 742538, 39%
%% Number of notices to a more than 10 domains = 483590, 26%
%% Number of notice received by recipients 
%% Number of notice per notice principal
%We identify \TotalNumberOfNoticeRecipient~notice recipients 

%\TotalNumberOfInfringingURLs~ URLs
% \TotalNumberOfNoticeSenders~unique senders
%We identify complaints from 38,523 unique senders, covering 1,054,248,823
We identify complaints from 38,523 unique senders, covering 1.05 Billion URLs, which are hosted across \TotalNumberOfDomains~domains. Individual complaints tend to contain multiple URLs, with an average of 560 URLs and 31 domains per notice. In terms of the  \emph{types} of notices, there is a remarkable skew towards DMCA complaints.
This is partly driven by the prominence of search engine recipients within our data. 
Table~\ref{tbl:dataset} presents a breakdown of the types of complaints across the entire dataset. We find that DMCA notices make up 98.6\%  (1.05B URLs) of the dataset and a similar share of domains (97.8\%).
DMCA notices are a US-enforceable complaint which covers takedown notices for (allegedly) copyright infringing content.
%Interestingly, while DMCA notices represent most of the notices, they only cover 40\% of the recipients. 
The senders largely appear to be third party organisations who act on behalf of the actual copyright owners: just 9\% of notice senders listed are also the principal. This contrasts with a 2006 study~\cite{Urban2005} where 98.5\% of notices were sent by right owners, suggesting a shift in behaviour (as seen in~\cite{Urban2016}).
Measured by frequency, DMCA notices are followed by Defamation (52K), Court Orders (29K) and Government Requests (2.7K), covering nearly a third of the recipients (31\%). We also observe a number of less popular complaint types, such as Law Enforcement Requests, Data Protection and Trademark infringements. Whereas these make up  less than 0.001\% of the dataset, they cover more than 20\% of the recipients.
%Despite of the staggering share of complaints done by third-party copyright organisations (DMCA notices), the divergence between the share of notices (by type) and the fraction of domains affected by it, points to a rich and diverse ecosystem where different practices exist. \ignacio{Fix now we have \% of domains, numbers have changed and the current story doesn't hold} 

%%GT: I removed the below paragraph as it felt like we might be over-egging the limitations a bit
%Note that it is impossible to exactly quantify the extent to which this distribution is driven by Lumen's dataset \vs the actual underlying complaint ecosystem. Hence, for the rest of the paper we emphasise that our insights are specific to the 1 billion complaints collected by Lumen. As Lumen includes some of the largest and most critical stakeholders of the Web we contend that the insights are highly valuable.
% (1.05B URLs in XXX domains complained about)
%just 9\% of notice senders listed are also the principal-: % with no listed principal?

\begin{table}[t]
\centering
\resizebox{\columnwidth}{!}{% 
\begin{tabular}{@{}llllll@{}}
\toprule
\textbf{\begin{tabular}[c]{@{}l@{}}Notice\\ Type\end{tabular}} & \textbf{\begin{tabular}[c]{@{}l@{}}\% of \\ notices\end{tabular}} & \textbf{\begin{tabular}[c]{@{}l@{}}\% of\\ URLs\end{tabular}} & \textbf{\begin{tabular}[c]{@{}l@{}}\% of\\ senders\end{tabular}} & \textbf{\begin{tabular}[c]{@{}l@{}}\% of\\ principals\end{tabular}} & \textbf{\begin{tabular}[c]{@{}l@{}}\% of\\ domains\end{tabular}} \\ \midrule
DMCA                                                           & 98.6                                                              & 99.99                                                         & 94.46                                                            & 99.87                                                               & 97.78                                                              \\
Defamation                                                     & 0.95                                                              & \textless 0.01                                                & 0.15                                                             & \textless 0.01                                                      & 1.50                                                              \\
Court Order                                                     & 0.19                                                              & \textless 0.01                                                & 4.87                                                             & 0.07                                                                & 0.41                                                               \\
Government Request                                              & 0.15                                                              & \textless 0.01                                                & 0.15                                                             & 0.02                                                                & 0.13                                                               \\

Private Information                                             & \textless 0.01                                                    & \textless 0.01                                                & 0.071                                                            & -                                                                   & 0.003                                                              \\
Data Protection                                                 & \textless 0.01                                                    & \textless 0.01                                                & -                                                                & -                                                                   & <0.001                                                               \\
Law Enforcement Request                                          & \textless 0.01                                                    & \textless 0.01                                                & 0.02                                                             & 0.03                                                                & <0.001                                                               \\
Trademark                                                      & \textless 0.01                                                    & \textless 0.01                                                & 0.02                                                             & -                                                                   & <0.001                                                               \\ 
Other                                                          & 0.08                                                              & \textless 0.01                                                & 0.26                                                             & \textless 0.01                                                      & 0.16                                                               \\ \hline
Total                                                          & 1,885,267                                                         & 1,054,248,823                                                 & 38,523                                                           & 20,686                                                              & \TotalNumberOfDomains                                                                \\ \bottomrule
\end{tabular}%
}
\caption{Dataset summary with the percentages of notice types, and the corresponding share of URLs, senders, principals and domains.}
% \ignacio{percentages and a line for totals? -cosmetics \dami{done}}
\label{tbl:dataset}
\end{table}

%%%%%%%%%%%%%%%%%%%%%%%%%%%%%%%%%%%%%%%%%%%
% \subsection{Characterising Notice Senders}
\subsection{Who Are the Notice Senders?}
\label{sub:who_injects}
%%%%%%%%%%%%%%%%%%%%%%%%%%%%%%%%%%%%%%%%%%%

%%% sec A
%% number of notice senders that that target fewer than 100 domains
%% Number of complaint made by notice senders complaining about more than 100 domains.
%% Number of distinct domains contributed by senders with more than 100 domains sent  1,052,045

%Percentage of notices issued against a single domain
%Number of notice senders that generate a single complaint
%Median number of notices for senders that send multiple notices repeatedly to a single domain
%mean number of notices for senders that send multiple notices repeatedly to a single domain
%percentage of notice senders that generate multiple complaints to multiple domains

%%%%%%%%%%%%%%%%%%
%number of senders that issue just a single complaint = 37%
%number of senders that issue more that a 100 complaint = 94%
%number of complainants that make a single compaints just once =75%
%distribution of complaints based on sender category

The previous section suggests that notice senders are more diverse than simply the owners of copyright material, and that different reporting practices coexist. 
To shed light on these aspects, we next inspect the entities behind the notices submitted. We find that the distribution of notices is \emph{highly} skewed towards a few extremely active senders. 
%\newt{This is also the case in the previous study by Urban \etal\cite{Urban2016}.} 
The top 10\% of notice senders report over 1 billion URLs, in stark contrast to just 550K by the bottom 90\%.

\begin{figure} [t]
\includegraphics[width= 8.5cm]{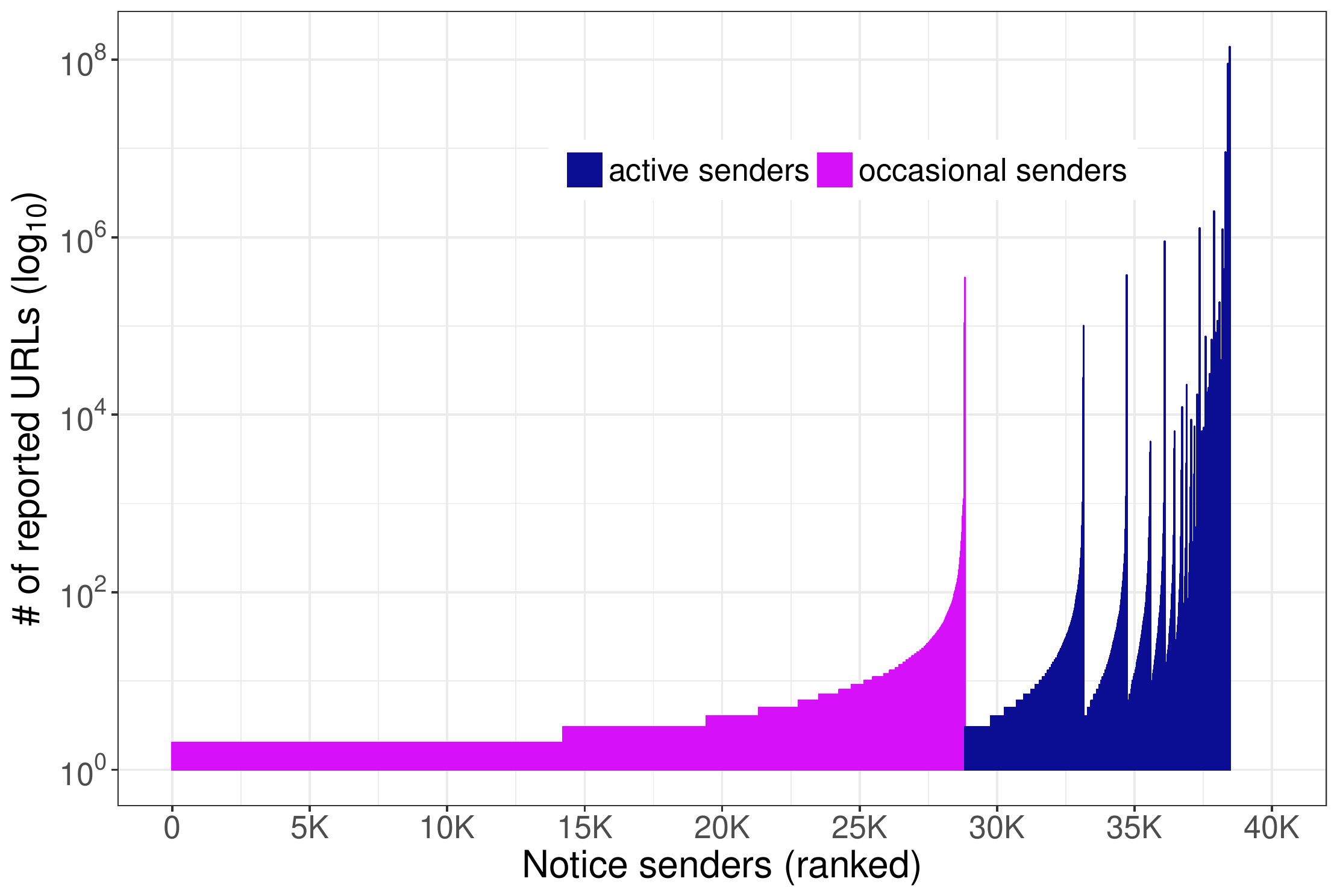}
\caption{Number of  reported URLs per complainant. The X-axis present each complainant, sorted by the two metrics: \one~number of days a complainant reports on; and \two~notices generated by each complainant.  \gareth{If possible update colours}
% --hence, the spikes. 
% with  \emph{occasional} senders, being those issue  notices on as single day.
}
\label{fig:hist_sender_urls}
\end{figure}

% \gareth{All purple senders only send on a single day, so what is the secondary ordering metric?\dami{Number of notices received}}

% Figure~\ref{fig:hist_sender_urls} presents how many notice senders generate a given volume of complained URLs.
% The distribution is  highly unbalanced, with 94\% of notice senders issuing fewer than 100 complaints and 37\% making only a single complaints.
% We find that the majority of notice senders are occasional users of the complaint system. 
% Thus, this analysis reveals two broad categories of complainants, highlighted in Figure~\ref{fig:hist_sender_urls}: \one~\emph{active}, and \two~\emph{occasional} notice senders, only generating notice(s) on a single day. That said, multiple notices can be sent on this single day and, in fact, we observe an average of XX notices generated by these occasional senders. 25\% of complainants are the former, injecting 99.9\% of all complaints that we observe. Even though the majority of complainants are \emph{occasional}, the smaller population of \emph{active} complainants still generate the bulk of the notices. 
% In total, occasional senders jointly contribute just 0.08\% of the total number of complaints made. 

Figure~\ref{fig:hist_sender_urls} presents the number of reported URLs from each notice sender. The X-axis is sorted by two metrics: \one~the number of days that a sender generates notices on; and \two~the number of notices sent (note that Y-axis is in log scale).  
The distribution is highly unbalanced, with a large majority  of notice senders (94\%) issuing fewer than 100 complaints.
The figure reveals two broad categories of complainants: \one~\emph{active}, who send complaints on multiple days and \two~\emph{occasional}, only generating notices on a single day.  
Whilst the \emph{active} group represents just 25\% of all complainants, it is responsible for almost all notices (99.92\%). In contrast, the \emph{occasional} senders, consisting of the remaining 75\%, collectively contribute just 0.08\% of the total number of complaints. 
It can also be observed that the curve in Figure~\ref{fig:hist_sender_urls}  is not monotonic. This is because the number of notices issued each day can vary.
Whereas occasional senders, by definition, \emph{only} issue complaints on a single date, some send multiple notices. Although the daily average is just 1.4 notices, there are some occasional senders who send a large number of complaints in a single burst. For example, Idreto (an occasional sender) generated 350K notices on a single day. As a result, even occasional senders can have a significant impact on the overall ranking of reported domains. 

We now inspect the types of complaints generated by these two groups of notice senders. Table~\ref{tbl:top_10_notice_senders} presents the top 10 complainants. 
Similar to the findings of \cite{Seng2014,Urban2016}, we find that \emph{active} senders are dominated by copyright enforcement and trade organisations \eg the British Phonographic Industry (BPI), Apdif Brasil, Apdif Mexico, \etc
These organisations represent hundreds of music recording companies in their respective jurisdictions. Also within this group are copyright protection agencies such as Muso, Aiplex, Mark Monitor and Entura who specialise in tracking pirated content. These companies aggregate complaints from many different copyright holders, and act as enforcers on their behalf. This partly explains their broad coverage and ability to produce large volumes of complaints. In contrast, \emph{occasional} senders are predominantly made up of small organisations and private entities. We see that the categories of \textit{Private Information} and \textit{Trademark} are particularly dominated by \emph{occasional} notice senders, making up 94\% and 70\% respectively. Conversely, \emph{active} senders hold sway among other notice types, \ie DMCA. These striking differences between the categories of sender highly differing strategies based on the nature of the complaints, with the ability of a small hub of complainants to dominate the wider system.

%Seemingly, despite the emergence of many different complaint types, large-scale copyright enforcers continue to dominate. 

% We further explore these two groups of senders by inspecting the number of complaints made based on the total number of days under observation for which they make complaints. Figure~\ref{fig:num_of_reported_urls_vs_num_of_days} plots the average number of reported URLs by notice senders that record complaints a total of $x$ days. It also plots the total number of complainants that make complaints for the same number of days (\ie the $red$ line). \dami{I have re-written it again, please check}
% \gareth{To be honest, it's still very cryptic. Also I think the fig needs a legend/label updates? Let's wait until/if the plot is regenerated befor rewriting (e.g. I think you were gonna make a box plot?). }

% \begin{figure} [t]
% \includegraphics[width= 8.5cm]{num_of_reported_urls_vs_num_of_days}
% \caption{The average number of complaints made by notice senders for the total $x$ number of days for which they make complaints along side the total number of complainants sending notices for that $x$ total number of days indicated by the $red$ line.}
% \label{fig:num_of_reported_urls_vs_num_of_days}
% \end{figure}

\begin{table}[]
\centering
\resizebox{\columnwidth}{!}{%
\begin{tabular}{@{}lllll@{}}
\toprule
\textbf{\begin{tabular}[c]{@{}l@{}}Notice\\ Sender\end{tabular}} & \textbf{\begin{tabular}[c]{@{}l@{}}\% of\\ reported URLs\end{tabular}} & \textbf{\begin{tabular}[c]{@{}l@{}}\% of \\ notice\end{tabular}} & \textbf{\begin{tabular}[c]{@{}l@{}}\% of\\ reported domains\end{tabular}} & \textbf{\begin{tabular}[c]{@{}l@{}}\# of \\ reported days\end{tabular}} \\ \midrule
Rivendell                                               & 13.17                                                         & 1.65                                                    & 4.97                                                             & 357                                                            \\
Aiplex                                                  & 9.76                                                          & 1.88                                                    & 1.30                                                             & 364                                                            \\
BPI                                                     & 8.60                                                          & 2.52                                                    & 2.72                                                             & 355                                                            \\
Apdif Mexico                                            & 8.52                                                          & 0.55                                                    & 0.16                                                             & 208                                                            \\
Mg Premium                                              & 7.77                                                          & 0.47                                                    & 0.49                                                             & 341                                                            \\
Apdif Brasil                                            & 7.56                                                          & 0.52                                                    & 0.39                                                             & 244                                                            \\
Remove Your Media                                       & 7.28                                                          & 0.92                                                    & 2.45                                                             & 346                                                            \\
Mark Monitor                                            & 5.29                                                          & 1.13                                                    & 5.12                                                             & 365                                                            \\
Fox Group Legal                                         & 4.36                                                          & 0.14                                                    & 2.26                                                             & 355                                                            \\
Protek Media, S.C.                                      & 4.20                                                          & 0.31                                                    & 0.67                                                             & 365                                                            \\ \midrule
Total                                                   & 806,358,505                                                   & 188,158                                                 & 224,076                                                          &                                                                \\ \bottomrule
\end{tabular}%
}
\caption{Top 10 complainants (by \# of reported URLs).}
\label{tbl:top_10_notice_senders}
\end{table}

\subsection{What Topics Do Complaints Target?}
\label{sub:topics}
%%%%%%%%%%%%%%%%%%%%%%%%%%%%%%%%%%%%%%%%%%%

\begin{figure} [t]
\includegraphics[width=8.5cm]{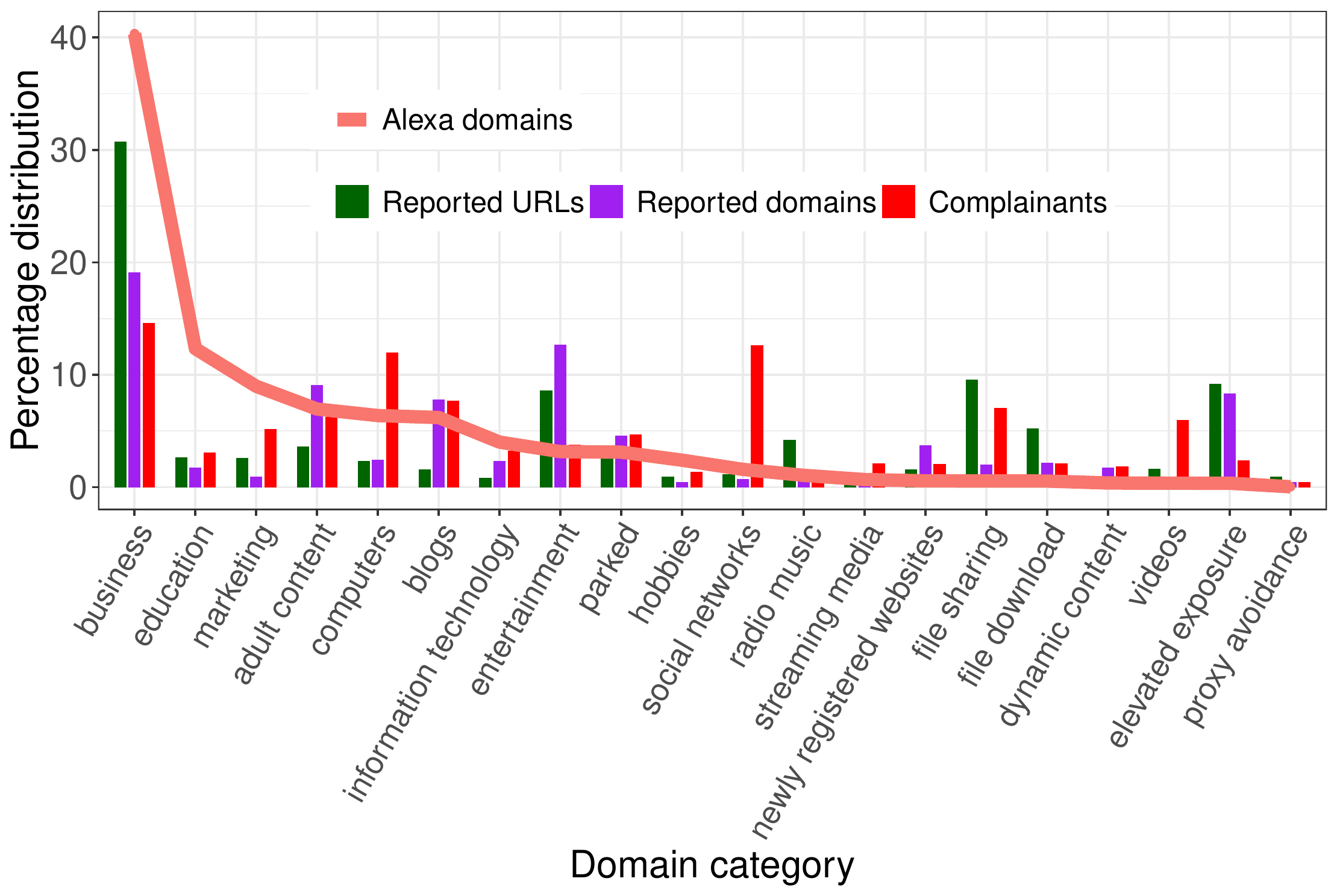}
\caption{Percentage of reported URLs, domains and complainants for the 20 most reported domain categories \vs the share of Alexa Top 100K  per category.}
\label{fig:domain_vt_classification_alexa}
\end{figure}

To better understand the main drivers behind the complaints, 
we investigate which categories of websites notices senders complain about.  
Figure~\ref{fig:domain_vt_classification_alexa} presents the percentage breakdown of complaints, domains and complainants   based on the reported domain category. Although it is clear that these classifications contain noise, we believe they offer useful insight into wider activity.  
To better understand how this relates to the general web ecosystem, we also depict the distribution of the Alexa Top 100K websites.

There is a  remarkable divergence between our dataset and Alexa, highlighting a clear bias towards complaints for certain types of websites. 
The largest fraction of domains are categorised as business (for both Alexa and Lumen), which includes websites such as \texttt{4shared.com}, \texttt{mangapark.me} and \texttt{gorillavid.in}.\footnote{We note that this category covers a number of business facing websites, including those engaged in Hollywood copyright theft.}
The most over-represented categories in the Lumen dataset are 
Elevated Exposure\footnote{This category refers to malicious websites that camouflage their true nature} (8\% \vs 0.3\% for Alexa),
Entertainment (13\% \vs 3\%),
File Download (2\% \vs 0.6\%), 
File Sharing (2\% \vs 0.6\%), 
Blogs (8\% \vs 6\%), 
and Adult Content (9\% \vs 7\%). 
While File downloading and sharing are expected, the other ones are less intuitive.
In contrast, categories that are under-represented include Education (2\% for Lumen \vs 12\% for Alexa) and Marketing (1\% \vs 9\%).

\begin{figure} [t]
\includegraphics[width=8.5cm]{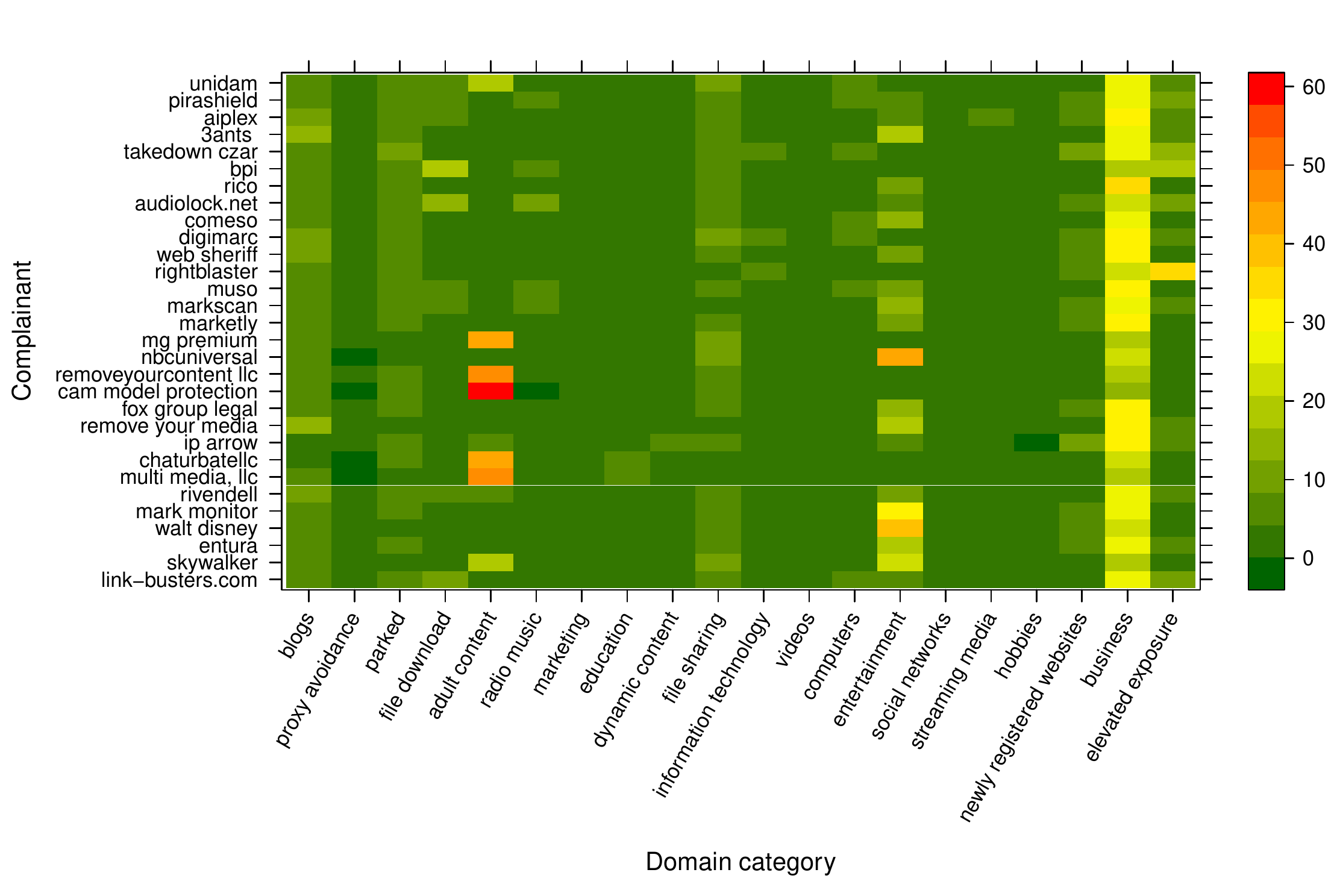}
\caption{Distribution of (top 20) domain categories (in \%) reported by (top 30) complainants. \gareth{Recompute based on how over represented domains are in terms of complaints vs overall population.\ignacio{Cosmetic:`complainant':`Complainant', `domain category': `Domain category'}}}
\label{fig:percent_dist_category}
\end{figure}

These trends suggest that notice senders tend to focus on individual categories. 
To explore this,  Figure~\ref{fig:percent_dist_category} presents the fraction of complaints that target each of these top 20 categories for the  top 30 complainants.
In-line with our conjecture, several complainants exhibit significant bias towards a single category. For example, 42\% of MG Premium's complaints target adult websites, 45\% of NBCUniversal's focus on Entertainment, and 34\% of Rico's is Business, demonstrating the prevalence of a high degree of specialisation.
Considering the practice of using web crawling to collect links~\cite{Urban2016}, this specialisation makes sense as it allows individual organisations to streamline their activities (based on which sites they have crawlers for). 

Briefly, we also note that these categories tend to attract different \emph{types} of notices too. For example, we find that the majority of Data Protection (60\%) notices are logged against adult domains. This aligns with the well known high litigiousness of the adult content industry~\cite{sag2014copyright}.
% In contrast, the bulk of the complaints made under Defamation and Other notice types pertain to search engines (22\% and 27\% respectively).
In contrast, 10\% of complaints catagorised as Defamation pertain to news. The remaining notice types (DMCA, Government, and Private Information) are typically made about business domains, representing 31\%, 33\% and 60\% for each category respectively; with the exception of Court Order notices where most complaints are regarding shopping domains (27\%). These clear trends confirm that the majority of notice senders are quite focused in their activities, with clear specialisation. 
%, while 13\% of Other goes to portal category.  
%Note that the business category broadly captures many commercial activities that do not fall squarely into the alternative categories mentioned above. 

%When combining the above observations, we confirm that complainants tend to target specific categories of websites, particularly those that are ripe for abuse, \eg File Sharing, Adult Content and Social Networks (due to the ability of users to upload content). This has clear societal implications, when attempting to understand the areas of the web that are most heavily abused; we later revisit these categories to understand the activities of their websites (\S\ref{sub:target_behaviour}).

% \gareth{What are these percentages specifically referring to, how are they calculated? Be careful when just putting percentages in parentheses, it needs to be very obvious what they mean. }
% \ignacio{any details/examples we can mention here?}

%%%%%%%%%%%%%%%%%%%%%%%%%%%%%%%%%%%%%%%%%%%%%%
%\section{Characterising Targets}
\section{Websites \& Webpages}
\label{sec:characterising_reported_domains}
%%%%%%%%%%%%%%%%%%%%%%%%%%%%%%%%%%%%%%%%%%%%%%

Whereas the previous section explored the complaints and complainants, we next inspect the websites (domains) and webpages (URLs) which the complaints pertain to. 
Specifically, we are interested in understanding which websites gain most attention, and their availability.

%%%%%%%%%%%%%%%%%%%%%%%%%%%%%%%%%%%%%%%%%%%
\subsection{How `Hot' Are Websites?}
\label{sub:hot_domains}
%%%%%%%%%%%%%%%%%%%%%%%%%%%%%%%%%%%%%%%%%%%

%% Percentage number of notices received by 1% of domains (63%)
%% Percentage number of notices received by top 100 domains (6%)
%% percentage of domains within alexa's top 100K  that receive less than 1K notice (95%)

%%% less ranked domain in alexa that receive more than 1000 complaints 2%

\begin{table*}[]
\resizebox{\linewidth}{!}{%
\begin{tabular}{@{}llllllll@{}}
\toprule
\textbf{Domain}           & \textbf{\begin{tabular}[c]{@{}l@{}}\# of \\ TLDs\end{tabular}} & \textbf{\begin{tabular}[c]{@{}l@{}}\# of times in \\ top 10\end{tabular}} & \textbf{\begin{tabular}[c]{@{}l@{}}Domain \\ category\end{tabular}} & \textbf{\begin{tabular}[c]{@{}l@{}}\# of \\ complainants\end{tabular}} & \textbf{\begin{tabular}[c]{@{}l@{}}Major complainant (s)\\ (\% of complaints)\end{tabular} }                                                                    & \textbf{\begin{tabular}[c]{@{}l@{}}\# of days \\ reported\end{tabular}} & \textbf{\begin{tabular}[c]{@{}l@{}}Alexa\\ Rank\end{tabular}} \\ \midrule
mp3toys.xyz      & 10                                                    & 3                                                                & elevated exposure                                            & 10                                                           & Apdif Brasil (99.9\%)                                                                                                            & 109                                                            & -                                                    \\
4shared.com      & 15                                                    & 4                                                                & filesharing                                                  & 40                                                           & Apdif Brasil (99.8\%)                                                                                                            & 246                                                            & -                                                    \\
googlevideo.com  & 1                                                     & 7                                                                & entertainment                                                & 84                                                           & Comeso (78.2\%), Remove Your Media (17.3\%)                                                                                      & 239                                                            & -                                                    \\
mangapark.me     & 3                                                     & 4                                                                & business                                                     & 18                                                           & Remove Your Media (99.1\%)                                                                                                       & 28                                                             & 3,901                                                \\
gorillavid.in    & 3                                                     & 3                                                                & business                                                     & 16                                                           & \begin{tabular}[c]{@{}l@{}}Fox Group Legal (55\%), Mark Monitor (25.2\%), \\ Vobile (17.1\%)\end{tabular}                        & 365                                                            & 6,778                                                \\
tvad.me          & 1                                                     & 2                                                                & business                                                     & 14                                                           & Fox Group Legal (59.9\%), Mark Monitor (39.9\%)                                                                                  & 94                                                             & 45,766                                               \\
israbox.vip      & 101                                                   & 5                                                                & media file download                                          & 19                                                           & Rivendell (99.6\%)                                                                                                               & 46                                                             & -                                                    \\
uploaded.net     & 3                                                     & 2                                                                & filesharing                                                  & 216                                                          & \begin{tabular}[c]{@{}l@{}}Rivendell (31.7\%), Skywalker(18.1\%)\\  Mark Monitor (14.9\%), Fox Group Legal (12.3\%)\end{tabular} & 365                                                            & 662                                                  \\
genteflowmp3.uno & 11                                                    & 2                                                                & media file download                                          & 6                                                            & Apdif Mexico (99.6\%)                                                                                                            & 149                                                            & -                                                    \\
deep-warez.org   & 1                                                     & 2                                                                & radiomusic                                                   & 43                                                           & Rivendell (98\%)                                                                                                                  & 345                                                            & 183,305                                              \\ \bottomrule
\end{tabular}%
}
\caption{Top 10 domains with most complaints and the number of TLDs associated with each domain, times it appeared in the top 10  (in terms of complaints) per month, domain category,  major complainant (with the share of complaints), number of days receiving complaints and its Alexa ranking position.}
\label{tbl:top_10_classification}
\end{table*}

%\gareth{Could we add Alexa rank? \dami{done}}

\begin{figure} [t]
\includegraphics[width= 8.5cm]{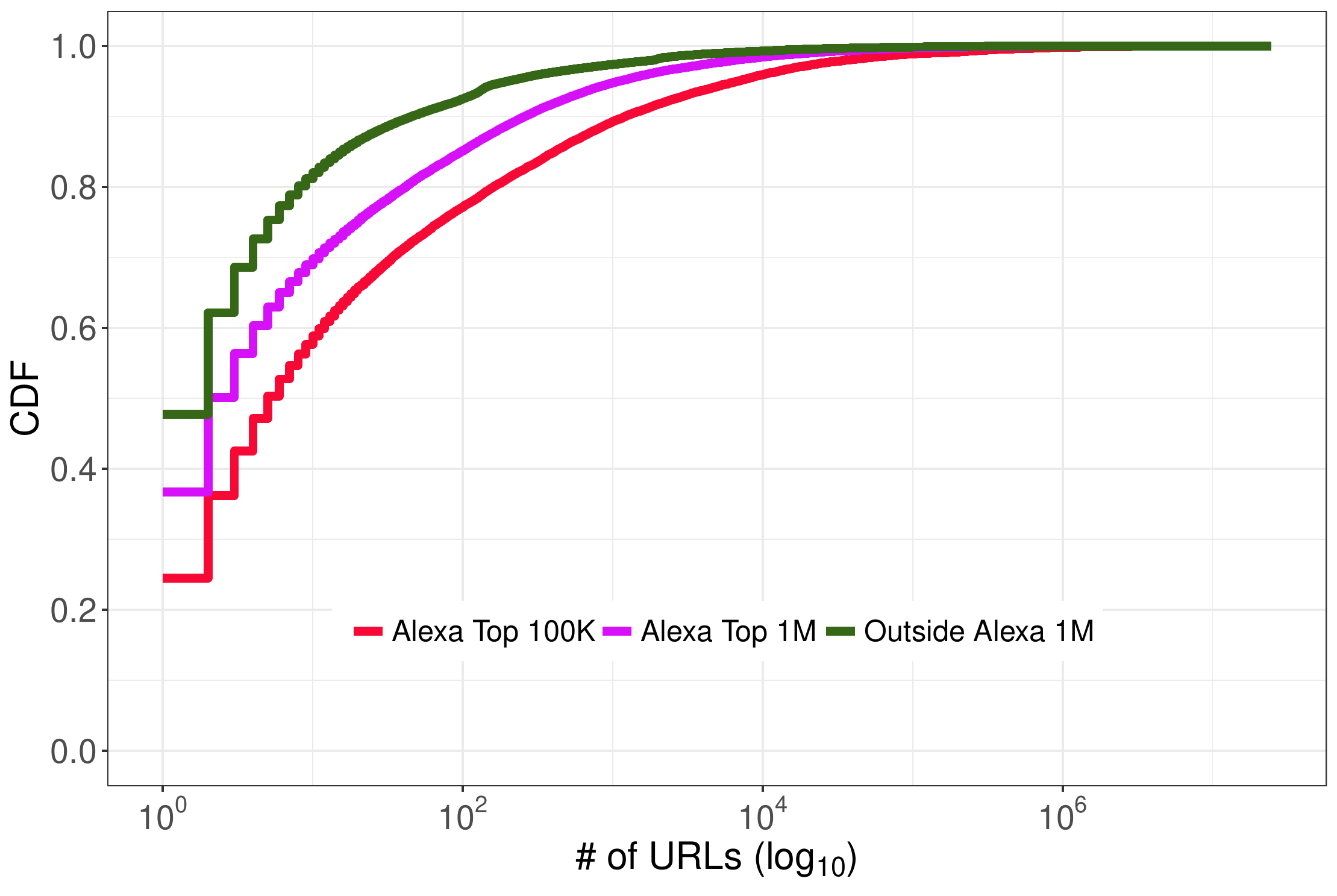}
\caption{CDF of the number of complaints per domain. Domains are classified by their Alexa Rank.}
\label{fig:cdf_alexa_urls_dist}
\end{figure}

\begin{figure} [t]
\includegraphics[width = 8.5cm]{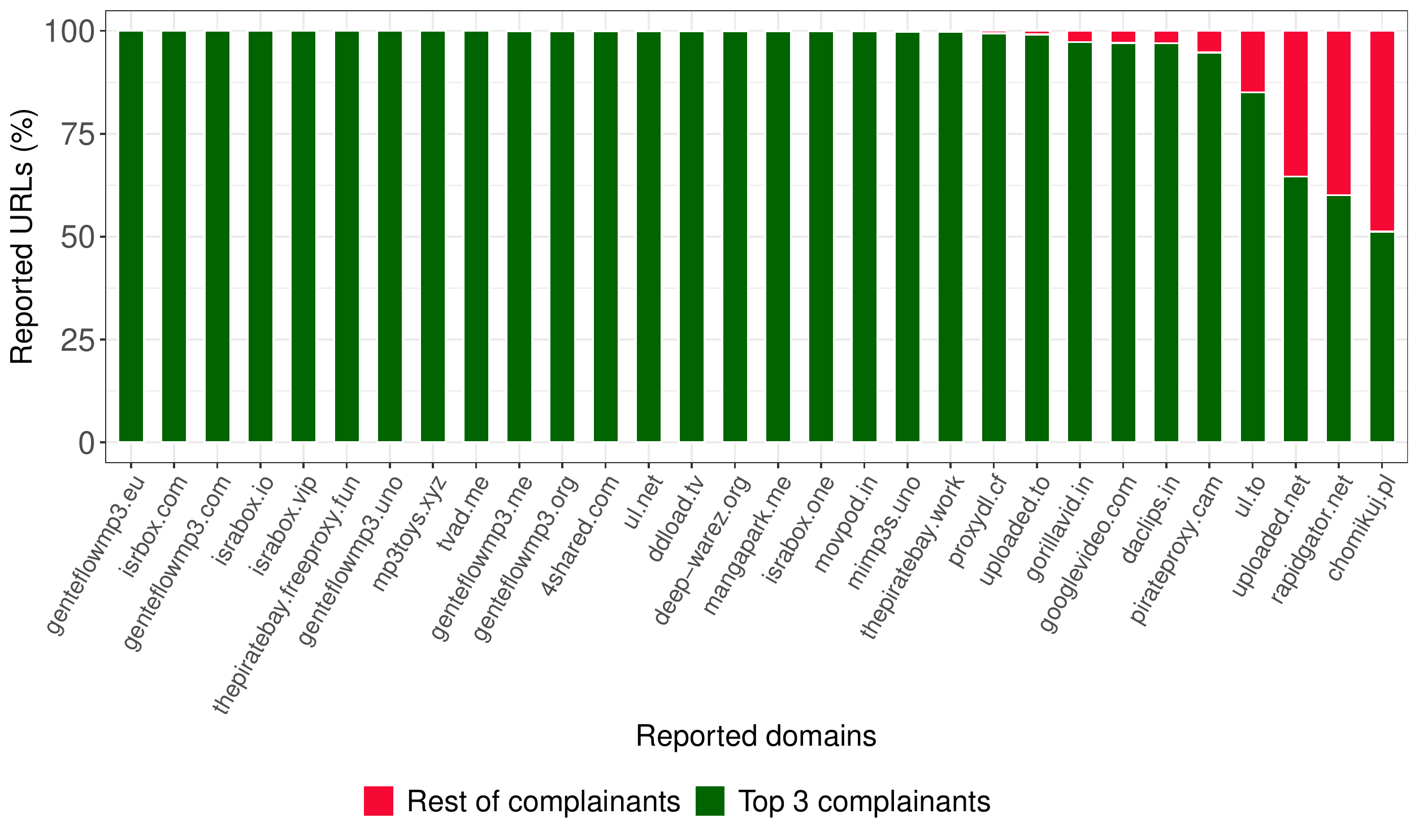}
\caption{Percentage of reported URLs by the top 3 complainants for each domain (on X-axis) \vs the rest of complainants.}
\label{fig:percent_dist_sender_domains}
\end{figure}

We start by inspecting the number of complaints generated about each website. Figure~\ref{fig:cdf_alexa_urls_dist} presents a CDF showing the distribution of notices per domain. We separate the data depending on the domain popularity, based on its position in the Alexa ranking.
Again, we observe a noticeable skew: it is clear that there are a small number of `hot' websites that gain the most attention from notice senders. In particular, the most reported domains (the top 1\% by the number of reported URLs), receive 63\% of all complaints.
%This mirrors our observations in \S\ref{sub:who_injects}, where we see a highly centralised set of notice senders. 
To provide insight into the characteristics of these domains, Table~\ref{tbl:top_10_classification} summarises the Top 10 in terms of the number of complaints. A range of uncommon websites are seen within this list. Although, overall, 60\% of notices relate to websites in the Alexa Top 100K, only 4 of the most reported 10 domains, and 8 out of the top 30 are within this Alexa rank (we find a Pearson  correlation of just 0.13 between these two rankings). 
The reasons behind these relatively obscure sites gaining `notoriety' are quite diverse, but they do share one common attribute: their position in the ranking is typically driven by a single complainant that repeatedly targets them. 
In fact, 9 out of the top 10 domains receive at least 50\% of their complaints from a single organisation. This is a general pattern across all domains: we find that 82\% of them receive at least half of their complaints from a single complaint. 
To visualise this dominance by a few complainants in each domain, we depict in 
Figure~\ref{fig:percent_dist_sender_domains} the percentage of URLs reported by the top 3 complainants of each domain (in green). The rest of complainants (in red) tend to contribute little to nothing. 
The vast majority of domains receive nearly all complaints from just a tiny set of senders. 
For example, \texttt{mp3toys.xyz} receives 99.9\% of complaints from a single party (Apdif Brasil).
Upon closer inspection, it is clear that this organisation uses an algorithm to `guess' potential infringing URLs based on song titles~\cite{Urban2016}. Although we envisage that these URLs are checked for liveness before complaints are generated, we note that \texttt{mp3toys.xyz} dynamically generates 200 OK HTTP responses for \emph{any} URL requested, likely disrupting any liveness checks.
These type of bulk sending activities do not appear to be rare occurrences. For example, the same organisation reported over 17M  URLs for file hosing domain \texttt{4shared.com}, despite the website only hosting 2M pages~\cite{torrentfreaknov2016}. As well as confirming that these highly active senders are heavily automated, it also suggests that rigorous procedures are not always followed. 

\subsection{Are Domains Unique?}

The above is based on unique domain counts, however, we also posit that some of these domains may actually host the same content, or even resolve to the same IP address. To test this, we turn to our DNS and webpage probes, which downloaded the HTML from all domains. We extract their \texttt{<title>} and all metadata tags. In cases, where two domains' tags match, we assume they host the same page. We term these \emph{replicas}.
The vast majority of domains host different content. 
Fewer than 0.01\% of domains have any replicas. 
From those that do, 73\% refer to the same IP address, indicating that the web server operator has simply created multiple domain names. Just 2\% of these have matching second-level domains (with a different TLD), whereas the remainder actually are entirely distinct. We conjecture that this may be an evasion tactic to avoid DNS-based blocking schemes. 
We also observe certain outliers; in the most extreme case, we find that \texttt{1fichier.com} uses 3,838 different domain names, which map to 78 IP addresses. This is a file sharing service well known for hosting illegal content. Another key driving force in the case of these extreme examples is the presence on numerous unblocker websites in our dataset. These websites essentially operate as proxies, generating third level domains for any website requested. For example, \texttt{s-s.www.cats.com.prx2.unblocksites.co} provides access to \texttt{www.cats.com} via \texttt{unblocksites.co}.
Table~\ref{tbl:top_20_second_level_domain} presents the top 10 (in terms of reported URLs) of these unblocking services. Remarkably, \texttt{unblocksites.co} actually constitutes 10\% of all reported URLs. In other words, we find that many complainants target these unblocker sites, by replicating their complaints for both the origin domain and the unblocked version. Understanding an exploring these is a key area of our future work.

\begin{table}[t]
\small
\centering
\resizebox{\columnwidth}{!}{%
\begin{tabular}{@{}lllll@{}}
\toprule
\textbf{\begin{tabular}[c]{@{}l@{}}Unblocking\\ Site\end{tabular}} & \textbf{\begin{tabular}[c]{@{}l@{}}\% of\\ Reported URLs\end{tabular}} & \textbf{\begin{tabular}[c]{@{}l@{}}\% of\\ Domains\end{tabular}} & \textbf{\begin{tabular}[c]{@{}l@{}}Unblocking Site \\ Category\end{tabular}} & \textbf{\begin{tabular}[c]{@{}l@{}}Alexa\\  Rank\end{tabular}} \\ \midrule
unblocksites.co                                                    & 10                                                                     & 0.13                                                             & uncategorised                                                         & 12,252                                                         \\
freeproxy.fun                                                      & 2.4                                                                    & 0.02                                                             & webproxy                                                              & 26,517                                                         \\
unblocked.lol                                                      & 2                                                                      & 0.04                                                             & proxy avoidance                                                       & 8,552                                                          \\
unblockall.xyz                                                     & 0.9                                                                    & 0.02                                                             & proxy avoidance                                                       & 820,291                                                        \\
proxydude.xyz                                                      & 0.6                                                                    & 0.007                                                            & elevated exposure                                                     & -                                                              \\
immunicity.gold                                                    & 0.5                                                                    & 0.03                                                             & proxy avoidance                                                       & -                                                              \\
unblockall.org                                                     & 0.5                                                                    & 0.01                                                             & business                                                              & 4,947                                                          \\
unblocked.cam                                                      & 0.5                                                                    & 0.02                                                             & proxy avoidance                                                       &                                                                \\
unblocker.cc                                                       & 0.4                                                                    & 0.009                                                            & proxy avoidance                                                       & 15,535                                                         \\
unlockproj.club                                                    & 0.4                                                                    & 0.02                                                             & uncategorised                                                         &                                                                \\ \bottomrule
\end{tabular}%
}
\caption{Top 10 unblocking services present in our dataset.}
\label{tbl:top_20_second_level_domain}
\end{table}

\subsection{How Stable are Domain Rankings? }
\label{sub:domain_stability}
%%%%%%%%%%%%%%%%%%%%%%%%%%%%%%%%%%%

%Median number of domains injected  that are within Alexa top 1M == 1,546

%Median number of domains injected per day across dataset = 11,802

%%% Number of domains complained about between 1 and 5 days
%% Number of domains complained about under 100 days
%% Number of domains complained about under 100 days and outside Alexa
%% Number of domains complained about over 300 days within ALEXA

%\begin{figure} [t]
%\includegraphics[width= 8.5cm]{top_10_domain_url_monthly}
%\caption{Top 10 domains with most URLs complained about each month. }
%\label{fig:top_10_domain_url_monthly}
%\end{figure}

\begin{figure} [t!]
\includegraphics[width= 8.5cm]{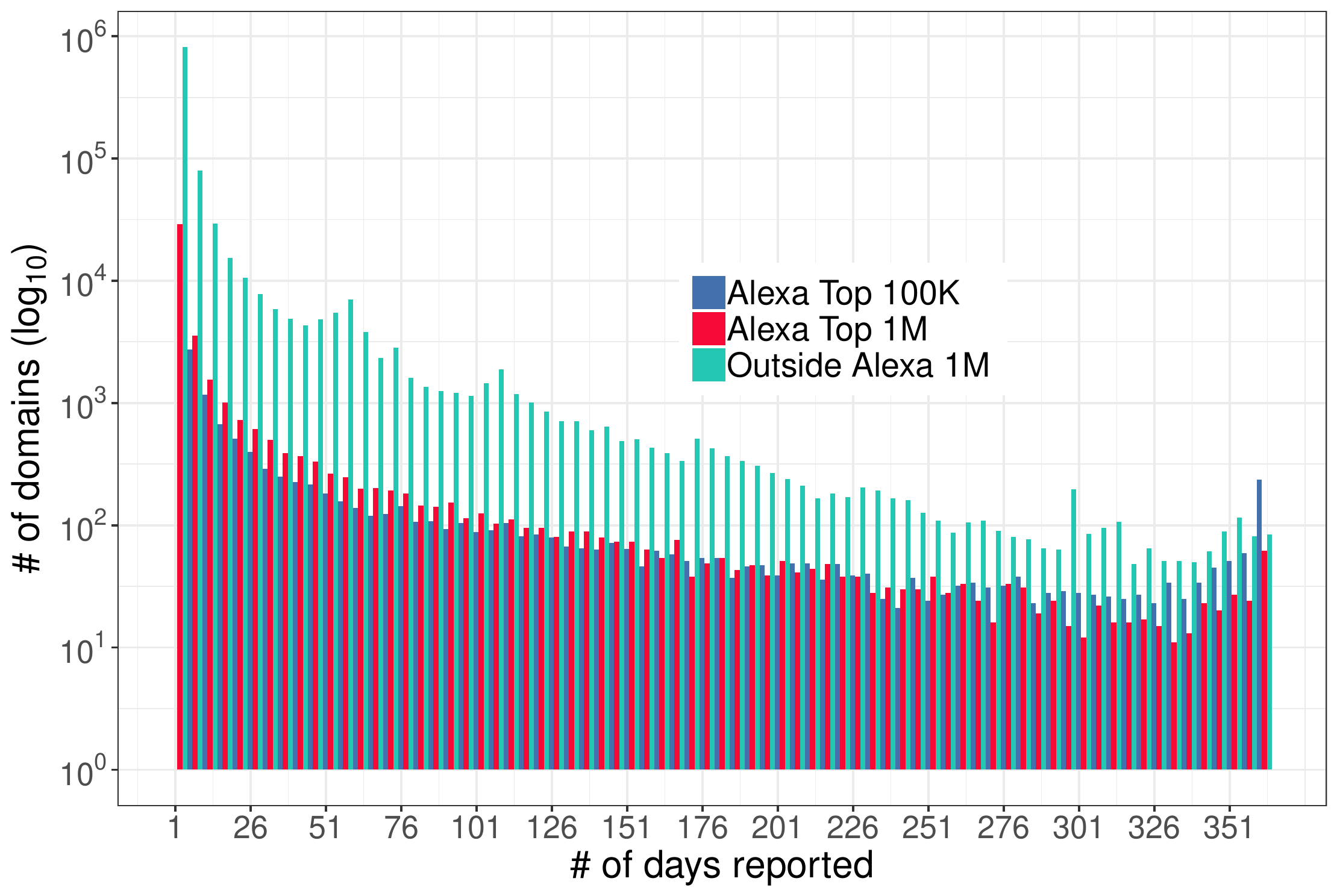}
\caption{Number of days receiving complaints for each domain in a given Alexa rank.}
\label{fig:number_of_domains_reported_days}
\end{figure}

The previous sections indicate that complaints about a domain are often dominated by a single sender. We conjecture that this dominance may result in significant temporal instability. 
This can be represented as a ranked list, capturing the domains most frequently reported on each day. 
To explore this, Figure~\ref{fig:number_of_domains_reported_days} presents the count of domains that are reported on $x$ days.
81\% of domains are complained about on fewer than 5 days, with only 0.2\% being complained about on more than 300 days. This suggests a high degree of instability, with the make-up of complaints changing on a daily basis. 
In line with our previous findings, the majority of these infrequently complained about domains are outside the Alexa Top 1M, whereas 87\% of domains which are complained about on over 300 days are within the top 1M. In Figure~\ref{fig:number_of_domains_reported_days} this can be seen as an upturn on the right-hand side of the graph. 

\begin{figure} [t!]
\includegraphics[width= 8.5cm]{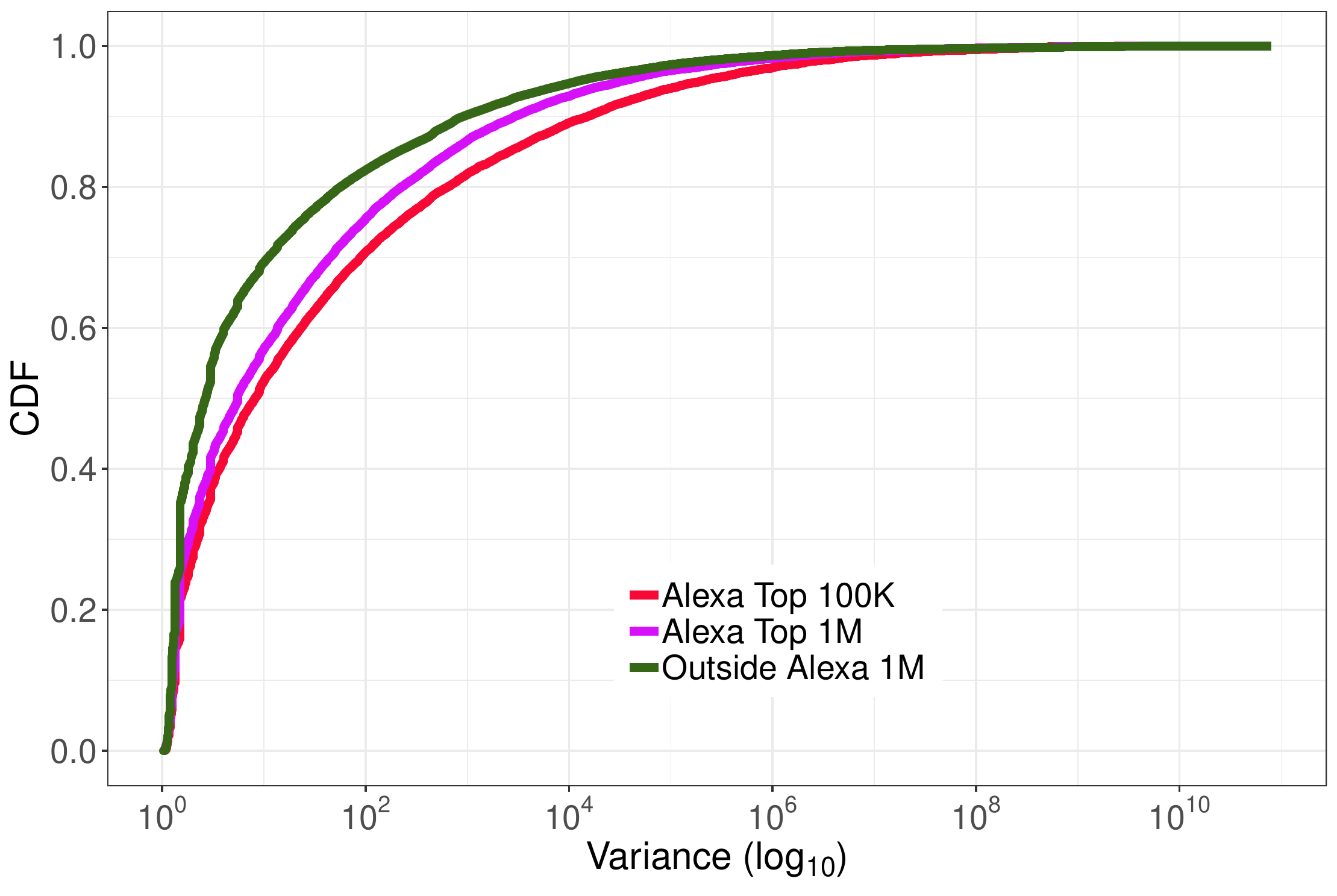}
\caption{CDF of the variance of the number of complaints per day for each domain in a given Alexa rank.}
\label{fig:cdf_variance_domain}
\end{figure}

To explore this trend, We next calculate the statistical variance of each domain's daily complaint count to understand how the number of daily complaints change. 
% \gareth{I guess we filter all domains with just 1 day of complaints?\dami{Yes}}
Figure~\ref{fig:cdf_variance_domain} presents a CDF showing the per domain variance over the entire dataset (for domains with complaints on multiple days). About 15\% of domains have a variance of zero because they receive the same number of complaints each day they are reported; 94\% of these are reported under six times.   
% \gareth{Am I right in guessing these are just domains which have 2 days of complaints? And I'm guessing it's just a small number of complaints? We could do with some context here for the 15\%.\dami{Well, Not all maximum we got is 34}  }
%Percentage of domains with variance of 0 reported only greater less than 5 times in the entire year
However, many remaining domains exhibit significant daily variance: 7\% of domains have a variance greater than 1K and, remarkably, 0.9\% of domains even have a daily variance greater than 1M.
This means that the number of complaints to a domain varies heavily on a day-to-day basis.

\begin{figure} [t!]
\includegraphics[width=8.5cm]{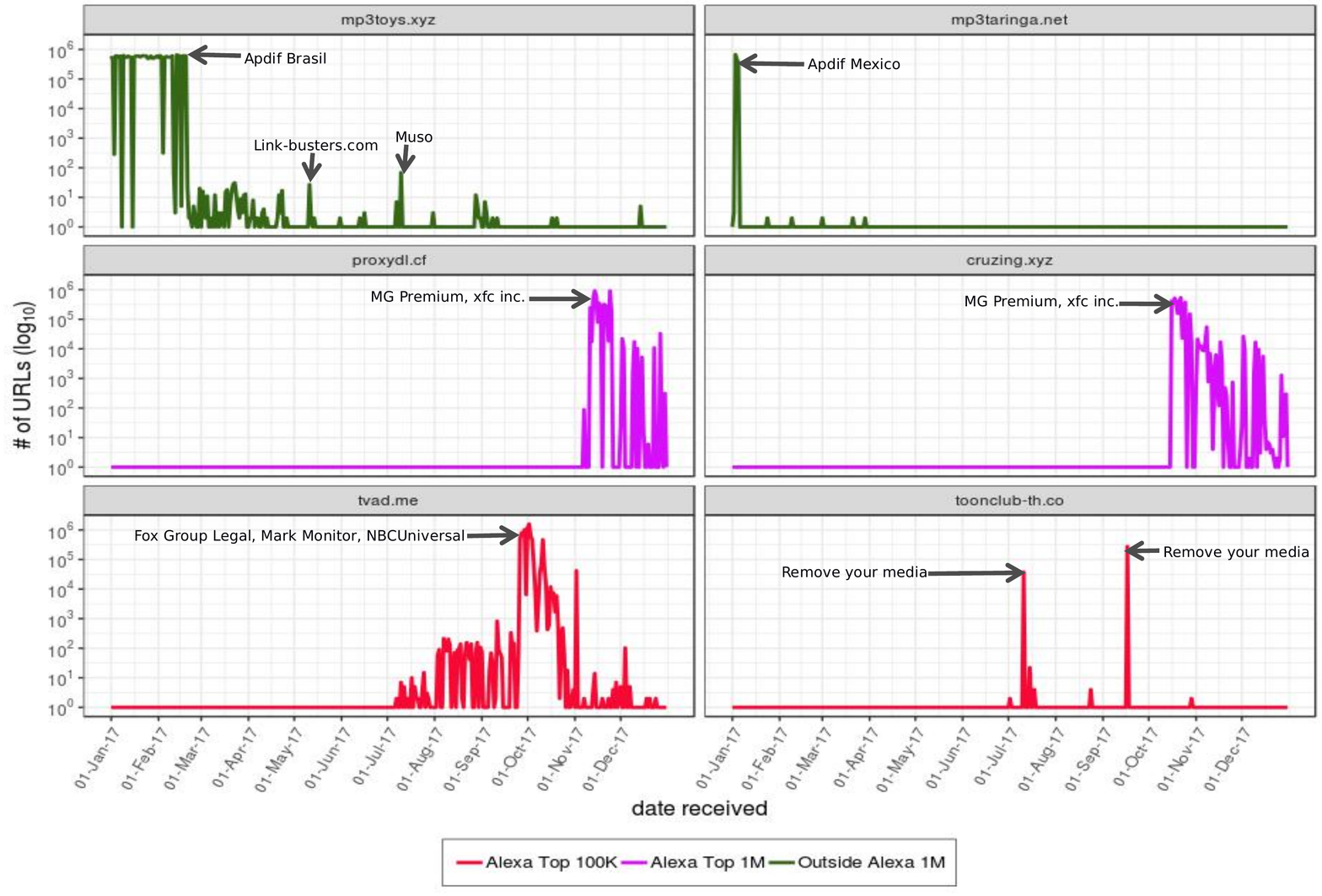}
\caption{Timeseries with complaints to selected domains and complainants causing complaints' bursts.}
\label{fig:timeseries}
\end{figure}

Closer inspection reveals that this is driven by \emph{extremely} aggressive complainants who periodically inject large sets of (sometimes repeat) notices.
For example, for domains with variance greater than 1M, we find that 92\% receive an average of 116 duplicates from the same sender. % Amongst the above 0.9\% of domains, XX\% of their complaints are generated by a single notice senders. 
To better highlight this aggressive activity, Figure~\ref{fig:timeseries} presents timeseries measuring the number of daily complaints for several example domains. We select the top two domains with the largest variance in each Alexa ranking category  (top 100K, top 1M, outside Alexa). Significant instability exists across each of these domains, with labels in the figure highlighting the individual sender causing each spike.
For instance, the top ranked domain, \texttt{mp3toys.xyz}, receives an average of 500K daily complaints from Apdif Brasil between Jan \& Feb, yet this collapses to below 6 per day in March. Similarly, \texttt{mp3taringa.net} receives a significant number of complaints between 03/Jan -- 05/Jan by Apdif Mexico, but following this the numbers collapse to  almost zero. 
We posit that these senders are not always careful in the complaints they generate but, rather, send bulk notices, leaving the recipient to make sense of the content.

%Finally, we compute the number of \emph{repeat} notices generated by senders, \ie notices that are repeatedly issued to a recipient with the same URL.
%This would suggest that the automated systems do not undertake detailed checks. We find a huge amount of repeat notices. When inspecting the 0.9\% of domains with variance greater than 1M, we find that 92\% receive duplicates from the same sender. On average, each sender generates 116 duplicates. This either indicates that complainants are overly aggressive in their behaviour, or recipients are not particularly reactive. That said, if the latter were true, it would be likely that the duplicates are spread uniformly across time (\ie sent as reminders). 

%An obvious conjecture is that some of these senders are generated this bulk of notices without properly verifying them, \ie so-called copyright trolling~\cite{sag2014copyright}.

% However, we find that XX\% of the duplicates are actually sent on the same day, with XX\% sent on consecutive days, suggesting this is not the case. 

\subsection{Are Reported Websites Alive?}
\label{sub:target_behaviour}
%%%%%%%%%%%%%%%%%%%%%%%%%%%%%%%%%%%%
%The previous sections have shown that there are a small set of complainants that focus the majority of their efforts on a relatively limited body of domains.

%We conjecture that the above model of complaint generation might result in certain evasion tactics emerging: a cat and mouse game. Although it is impossible to truly infer intent behind certain actions, we can observe trends. We next inspect the TLDs used by domains, how replicas are exploited and whether domains remove content. 

We finally inspect the availability of reported domains and webpage resources.  
As it is impossible to draw causal links between a notice being issued and the removal of content,  we limit our analysis to inspecting the availability of URLs, rather than inferring the reason for their (un)availability.
% It is impossible to draw causal links between a notice being issued and the removal of content, as we have no evidence that a given domain has received a complain or read it. Hence, we limit our analysis to inspecting the availability of URLs, rather than inferring the reason for their (un)availability.

\begin{figure} [t]
\includegraphics[width=8cm]{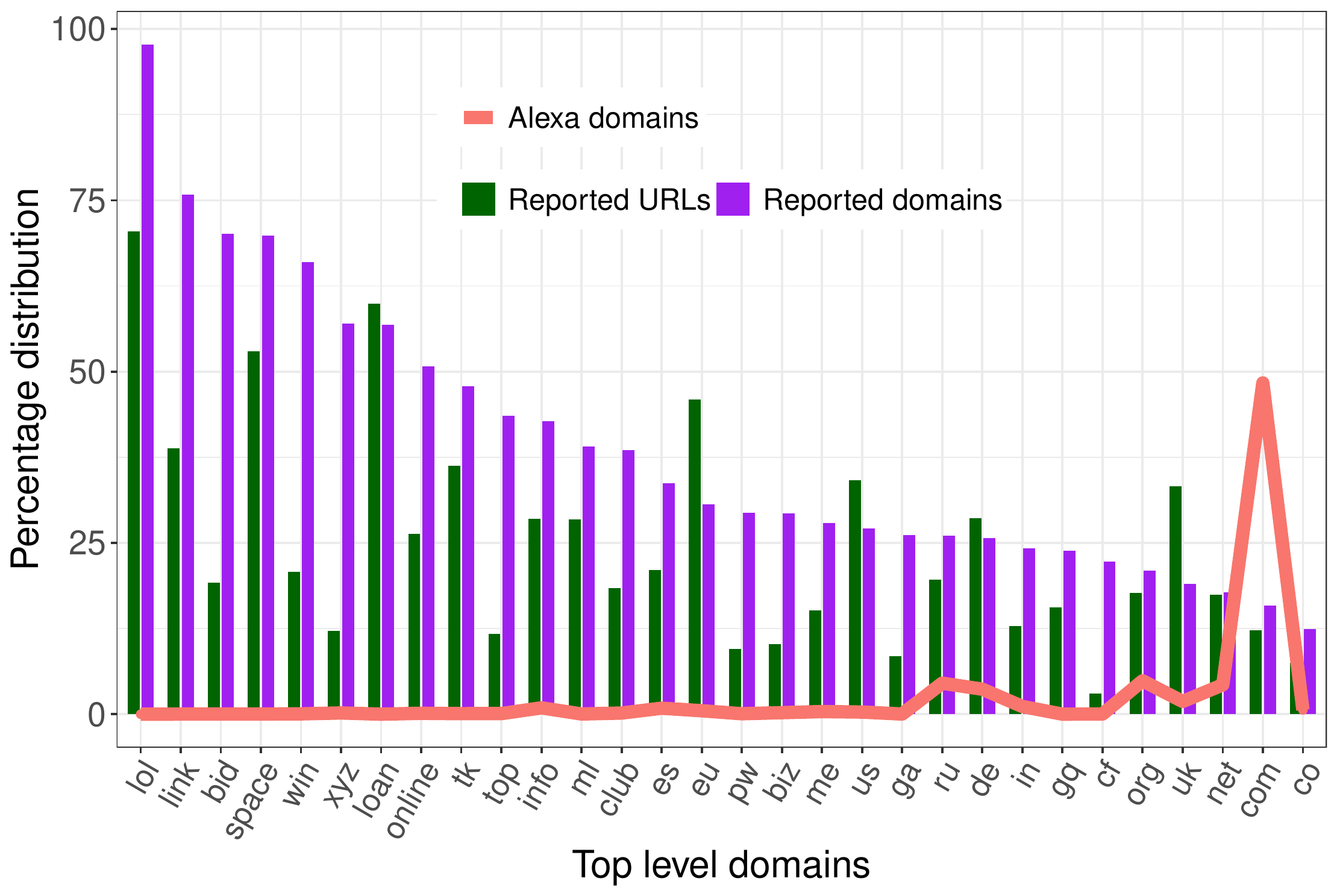}
\caption{Unavailable domains per TLD and URLs thus becoming inaccessible (as a \% from the total) \vs the share of Alexa domains for each TLD.}
\label{fig:nx_tld_urls_domain_alexa_comparison}
\end{figure}

%%%%%%%%%%%%%%%%%%%%%%%%%%%%%%%%%%%
\vspace{4pt}
\noindent\textbf{Understanding Domain Liveness.}
%%%%%%%%%%%%%%%%%%%%%%%%%%%%%%%%%%
We first check the liveness of each domain's DNS record using our DNS dataset. This reveals that 22\% of reported domains return an NXDOMAIN response.\footnote{This is returned when a domain name does not exist on the authoritative name server any longer.} 
These domains account for over 183M (17\%) of infringing URLs, with just 3\% (1,426) of them belonging to domains that rank within Alexa's Top 1M. 
As different domain registrars may have differing policies regarding the removal of records, we next group domains by their TLD, and check the likelihood of domains being taken offline; Figure~\ref{fig:nx_tld_urls_domain_alexa_comparison} presents the results. We plot both the number of websites (domains) that are unavailable, as well as the number of specific URLs that become unavailable (because the domain is offline). For context, we plot the density of Alexa top 1M domains that have each TLD. 
The majority of TLDs with a high percentage of NXDOMAIN responses do not frequently occur in the top Alexa rankings. 
Instead, we see that the majority of domains are from the recent wave of new generic TLDs~\cite{spencer2014much}. 
The most extreme is \texttt{.lol}, where 98\% of domains are NX; it is noteworthy that this TLD is operated by Uniregistry, which has been accused of predominantly hosting spam~\cite{lol_domain}. These trends indicate that the usage and behaviours across these TLDs are quite different, with some far more likely to contain unavailable domains. 

%Furthermore, this is driven by the large amount of domain parking we see amongst these new generic TLDs. For example, the top 8 TLDs all have between 50 and 62\% of their domains parked.\footnote{\url{https://ntldstats.com/}}

\vspace{4pt}
\noindent\textbf{Understanding URL Availability. }
Next, we turn to our periodic HTTP liveness checks to see if resources (URLs) are still alive (for those domains that do not return NXDOMAIN).
%Figure~\ref{fig:breakdown_repeated_check} presents the percentage breakdown of response codes 4XX, 5XX and timeout received from HTTP checks made to URLs. 
We see that the number of 200 (OK) responses decline slowly but steadily over the 4 week period that we monitored. After week 4, 22\% of URLs are inactive (\ie non-200). This trend, however, is relatively shallow with the majority of URLs (19\%) becoming inactive in the first week after the notice has been observed in Lumen.\footnote{Note that it is also possible that the URL was not live when the complaint was generated. Unfortunately, we cannot check this.} 
%Interestingly, the responses after a URL has become unavailable also differ. 
We also note that the statuses returned evolve across the four weeks. 
In week 1, the number of HTTP 4XX responses is 168K, yet in week 2 we only observe a further 12K webpages responding with 4XX. 
Instead, the number of HTTP 5XX and TCP timeouts increase, suggesting that these websites go through several stages that start with removing of content (therefore returning a 4XX) before total shutdown of the web server.  The latter makes sense, as it may be unnecessary to continue running a server if most content has been removed.

\begin{figure} [t]
\includegraphics[width= 8.5cm]{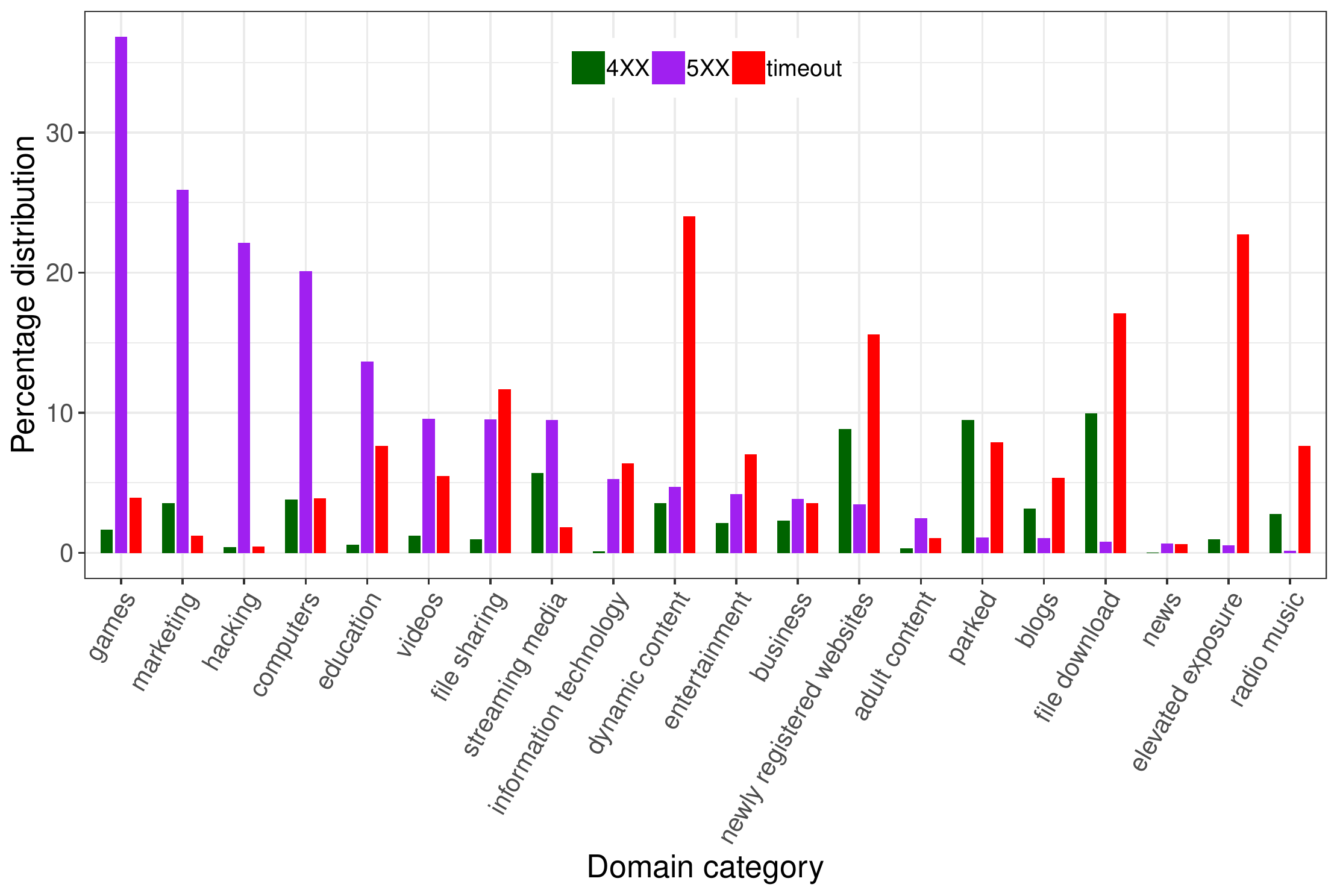}
\caption{URLs going offline (\ie non-$2XX$ response) for the top-20 most reported domain categories (in \%).}
% Non-$2XX$ responses of URLs, classified by category of domain

% Complaints responses (in \%, excluding $2XX$) for the 20 domain categories with most reported URLs.}
\label{fig:domain_category_not_response_200}
\end{figure}

%%%%%%%%%%%%%%%%%%%
\vspace{4pt}
\noindent\textbf{Understanding Category Availability}.
\label{sub:category_availability}
%%%%%%%%%%%%%%%%%%%%
We continue our analysis by inspecting which category of URLs are most likely to go offline. Figure~\ref{fig:domain_category_not_response_200} breaks down all URLs into their categories and presents the share within each group that reports non-200 HTTP responses after complaints are generated. Certain categories are significantly more likely to return non-200 responses. 
%For example, 37\% of URLs on  domains classified as Games return a non-200 response within 4 weeks of a complaint. 
For example, 10\% of URLs classified as File Download return a 404 response; similar traits are also seen across Parked (9\%) and Newly Registered Website (9\%).
In contrast, 24\% of URLs classified as \textit{Dynamic Content} (\ie websites that generate different material for each visit) return a TCP connection timeout, \ie the web server is no longer online. 
Other examples of categories with high numbers of URLs that timeout include \textit{Elevated Exposure} (22\%), and \textit{File Download} (17\%). 
The category which contains fewest unavailable sites is \textit{News}, where only 1.34\% of URLs become non-200. This indicates that the robustness of sites differs substantially across categories. That said, it is reasonable that domains more clearly engaging in suspicious activity are most likely to become unavailable. 
%We find a Person Correlation between number of reported URLs that are still available and dead across each domain to be 0.04. \gareth{How was this stat computed? What two columns were you correlating?\dami{this was computed by counting the number of 2XX and non 2XX for each domain reported and find the Pearson Correlation across.}}\gareth{Not sure what this tells us?}

% Those domains which become unavailable receive a median of XXX complaints \vs XXX for those who remain available during our monitoring period. 

% \gareth{What's the correlation between number of complaints and takedowns?}

%%%%%%%%%%%%%%%%%%%%%%%%%%%%%%%%%%
%\subsubsection{Understanding TLD Availability}
%%%%%%%%%%%%%%%%%%%%%%%%%%%%%%%%%

\begin{figure} [t]
\includegraphics[width=8.5cm]{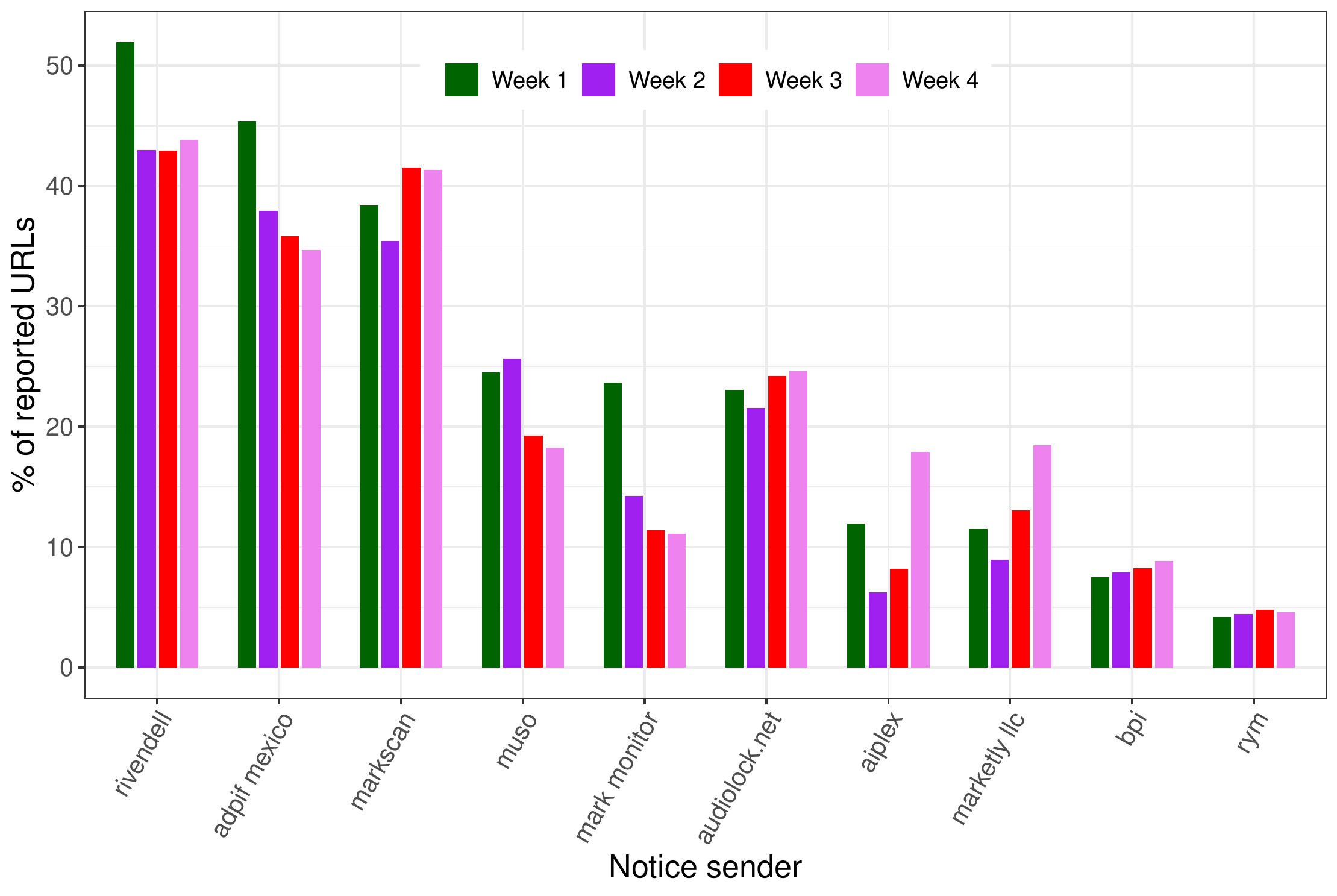}
\caption{Weekly share (non-cumulative) of URLs with a non $2XX$ response for the top 10 complainants.
% Each weekly bar excludes results in the previous week, \eg if a URL returned 404 in week 1, it is not included in the calculation of week 2.
\ignacio{different picture if done by domain? ie, a single over-reported domain being very reactive might bias this picture}
}
\label{fig:notice_sender_liveness_all_response_except_200}
\end{figure}

%%%%%%%%%%%%%%%%%%%
\vspace{4pt}
\noindent\textbf{Understanding Complainant Success}
\label{sub:complainant_success}
%%%%%%%%%%%%%%%%%%%%
An obvious question is which complainants are most likely to see their reported URLs deleted. 
To measure this, we calculate the percentage of reported URLs from each notice sender that later sees the URL resource returning a non-200 response. 
Figure~\ref{fig:notice_sender_liveness_all_response_except_200} presents the weekly percentage of complaints that return a non-200 response after  each weekly liveness check for the top 10 senders (based on the complainant with most not 2XX responses).
%In the 4 week checks performed on complaints, 97\% of notice senders have most failed checks as 4XX. 
The websites targeted by these different notice senders have very different availability properties. 
Complaints from rights enforcement organisations (\eg Rivendell, MarkScan, AudioLock.net) appear more effective compared to trade organisations (\eg British Phonographic Organisation). For example, about 53\% of complaints in the first week of submission from Rivendell return non 2XX response code whilst just 8\% from British Phonographic Organisation return same. These results confirms that the efficacy of these different organisations differs greatly, and that the websites they pursue have extremely different characteristics in terms of resilience to takedown. 

%that either the individual senders employ different tactics or, alternatively, the websites they target follow different strategies themselves. 

%\gareth{Minor point: FOr some reason, we always seem to write the complainants in all lower case. We should try to adhere to their proper names, e.g. LeakID rather than leakid}

%\gareth{COudl we compare occasionally vs. highly active senders}

% \begin{figure} [t]
% \includegraphics[width= 8.5cm]{notice_sender_week_liveness}
% \caption{Weekly percentage of URLs that return $4XX$ response code for the top 10 senders. Each weekly bar excludes results in the previous week, \eg if a URL returned 404 in week 1, it is not included in the calculation of week 2.}
% \label{fig:notice_sender_week_liveness}
% \end{figure}

%%%%%%%% all non 200 responses

\gareth{Possible thing to add if we have time: VTscore stuff}

%%%%%%%%%%%%%%%%%%%%%%%%%%%%%%%%%%%%%%%%%
%\section{Discussion}
%%%%%%%%%%%%%%%%%%%%%%%%%%%%%%%%%%%%%%%%%

%%%%%%%%%%%%%%%%%%%%%%%%%%%%%%%%%%%%%%%%%
\section{Related Work}
%%%%%%%%%%%%%%%%%%%%%%%%%%%%%%%%%%%%%%%%%

There are two main areas of related work: \one~studies that rely on Lumen for exploring web complaints; and \two~studies that have explored illegal online activities  more generally.
\vspace{4pt}

\noindent\textbf{Web Complaint Studies.}
There have been numerous studies into web complaints. Of most relevance are the small set of data-driven papers that have empirically explored the nature of takedown requests in the legal domain. 
For example, Heins \etal~\cite{Heins2005} examined 320 takedown notices to determine if the takedown process is fairly used by reporting organisations. They discovered that 20\% of notice senders send weak copyright claims, an assertion our results agree with. 
Previous studies from Seng~\cite{Seng2014} and Urban \etal~\cite{Urban2016} explored the entities behind notices from approximately 500K and 300K complaints, respectively. Both studies reveal that a small fraction of senders (mainly copyright protection companies and trade association) are responsible for the majority of notices. This is a result that is consistent with our findings. \cite{Urban2016} also classified the websites present in complaints; our work further builds on this to explore which categories of websites each sending entity specialises in.
In an earlier study, Urban \etal~\cite{Urban2005} examined nearly 900 notices in an attempt to discover the primary reporting organisations that file them. They also saw that business entities and corporations were the main users of takedown notices. To the best of our knowledge, our study is the largest of its kind. 
%Our work differs from these studies. Whereas they study a small sample of notices, we offer far broader vantage into web complaints, characterising the activities across notices regarding over 1 billion URLs. 
%Furthermore, we do not investigate the underlying motives behind complaints being generated but, rather, focus on key macro-level properties (\eg distribution of complaints across domains) and micro-level properties (\eg if a website removes the URL).

Closely related is the Right To Be Forgotten (RTBF), established in 2014 within the European Union. This allows individuals to request the de-listing of personal information from search engines, which is ``inaccurate, inadequate, irrelevant or excessive". Bertram \etal~\cite{bertramthree} inspected the RTBF requests issued to Google. This work is complementary to our own, as we focus on different complaint mechanisms, \ie Lumen does not cover RTBF. This is evident from the contrast between types of URLs in \cite{bertramthree} \vs our study, \eg 33\% of RTBF URLs relate to social media, and 20\% to news. This can be compared against \S\ref{sub:topics}, where we find a greater propensity towards Entertainment and Business URLs (driven by the prominence of copyright enforcement notices).

%We briefly note that we have utilised Lumen without our own past research~\cite{Ibosiola2018}, although this was limited to inspecting the complaints sent towards 33 cyberlockers. 

\vspace{4pt}
\noindent\textbf{Illegal Web Activities.} 
George \etal~\cite{George2007} examined the challenges that comes with the ease of sharing User Generated Content (UGC), highlighting the roles played by hosting providers and proposing stronger legislation to address the illegal sharing of UGC.  Wong~\cite{Wong2009} suggested a more flexible legislation, whilst Sawyer~\cite{Sawyer2009} recommended that platforms that share UGC should develop solutions to mitigate against the sharing of infringing material. Clay and Lucas discussed how a UGC platform (YouTube) has been exploited for such purpose~\cite{clay2011blocking,hilderbrand2007youtube}. Raman \etal also identified pirated content being shared via Facebook Live~\cite{raman2018facebook}.
To prevent the sharing of infringing content on such platforms, Dutta \etal proposed a signature-based detection to mitigate against infringing material remaining accessible online~\cite{dutta2008detecting}.

Peer-to-Peer networks and file hosters are also a frequently used to disseminate illicit or illegal material. Despite several anti-piracy efforts through the injection of fake content on BitTorrent portals~\cite{Cuevas2013,Cuevas2014,kaune2010unraveling,Farahbakhsh2013} and the shutdown of file hosters services~\cite{Lauinger2013b}, about 90\% of files shared using BitTorrent protocol are judged to be infringing~\cite{Watters2011,Robert2010}. Furthermore, 80\% of files shared through file hosters are also in the same category~\cite{Lauinger2013a}. Ibosiola \etal measured the availability of illegal content on streaming cyberlockers~\cite{Ibosiola2018}. They found that the majority of copyright infringing content is hosted on a small number of platforms. 
A large portion of complaints in our data also pertain to adult content. While there have been several studies on online adult content~\cite{tyson2013demystifying,tyson2015people,tyson2016measurements}, few focus on illegal adult content~\cite{hurley2013measurement}.
Of course, there are also a number of related studies looking at video content more generally~\cite{bottger2018open}.
Our work is orthogonal as we primarily focus on the complaints that are related to these activities. %Using a large dataset that provides us with a wide vantage into various types of complaints including both non-copyright issues and copyright complaints, we try to characterise the ecosystem at large. %We note that, even though Lumen contains notices regarding BitTorrent portals (\eg Pirate Bay), it does not specifically pertain to peer-to-peer. 

%%%%%%%%%%%%%%%%%%%%%%%%%%%%%%%%%%%%%%%%%%%%
\section{Discussion and Future Work}
\label{sec:conclusion}
%%%%%%%%%%%%%%%%%%%%%%%%%%%%%%%%%%%%%%%%%%%%

This paper has explored the nature of web complaints. With increasing scrutiny on illegal and illicit web activities, and the recent ability to streamline complaints against different stakeholders, this study offers a critical input into the wider ongoing debate about web governance and the use of so-called self-regulation~\cite{urist2006s}.

%understanding why, it is vital to understand why, how and which domains gain attention from the (little known) organisations discussed within the paper.

\vspace{4pt}
\noindent\textbf{Summary of Findings. } 
We have found a large and complex ecosystem dominated by a small set of complainants. While there are a large number of organisations (\TotalNumberOfNoticeSenders) that generate over 1 billion reported URLs, the top 10 complainants alone contribute over 41\% of all notices.
%The strategies of complainants range from very aggressive to extremely targeted, typically depending on the type of notices generated by the complainant. For example, copyright infringement complaints are spread far more indiscriminately than governmental notices. 
It therefore appears that these complaint procedures have become the dominion of a small group of large and very active organisations. Dominant players consists of a mix of influential copyright owners (\eg Fox) and third-parties specialised in pursuing copyright infringers (\eg Rivendell).
Bursts of complaints are common with most of the complaints towards each domain originating from 2--3 complainants, driving the unusual instability we see in the rankings.
Complainants are highly specialised in terms of the types of notices they generate and the domains they target. This leaves some domain categories (\eg File Sharing) regularly reported, and others rarely seen (\eg Education).
Surprisingly, many of the most frequently reported domains are quite obscure, and fail to score highly in popularity rankings.
Finally, we find that complaints do seem to matter. Many domain names are soon taken offline and 22\% of the URLs are inaccessible within just 4 weeks of us observing the complaints. Hence, it is clear that we shed light on a highly dynamic environment from the perspective of domain operators too.

\vspace{4pt}
\noindent\textbf{Societal and Legal Implications. }
Web governance and the (mis)use of web complaint mechanisms have important social and legal implications. Transparency is critical and, as a society, it is important to know how and why information is filtered. 
This is particularly the case as we have found that these mechanisms might not be always used wisely, \eg with some complainants generating hundreds of repeat notices, and seemingly auto-generated URLs  (\S~\ref{sub:domain_stability}). 
We argue that this might overwhelm recipients, who will not necessarily have the resources to deal with these large numbers.
%A further implication from our work is the observation that, despite being open to all, these complaint systems largely remain the preserve of a few large organisations. 
As these highly centralised models of operation have the potential for misuse, understanding the activity of senders is therefore critical. 

Our results further suggest that there is opportunity to improve and streamline the procedures. 
For notice recipients, the filtering of invalid complaints would no doubt be a valuable innovation. 
That said, we do not discount the veracity of many complaints, and  therefore developing mechanisms to support this process from the perspective of (notice senders would also be worthwhile. This, of course, should not be done at the expense of website operators, who should always be given paths to recourse.  Developing techniques that automate the above three things is important. 
Arguably, Lumen and similar platforms can play a powerful role in this process.

%Both domains hosting the content and complainers filing notices against it can potentially abuse the complain complaints mechanisms. We notice for instance that a significant number  of domains engage in elusive manoeuvres (\eg by creating replicas) while some complainers seem to complain indiscriminately with the hope that some of the complaints are effectively targeted and cause a reaction.

\vspace{4pt}
\noindent\textbf{Future Work. }
There are a number of lines of future work. 
First, we hope to expand our access to more diverse datasets. Within the paper, we have not investigated the role that recipients might play within the nature of complaints observed. This is likely to open up new lines of interesting investigation. 
%First, we emphasise that the Lumen dataset does not contain complaints made \emph{to} websites but, rather, \emph{about} websites, \eg requesting that Google remove a certain URL from their search results. Hence, we cannot definitively state that a website owner reacts to or ignores notices. 
%In our  future  work we will  explore exactly the types of notices that are issued to websites, and  capture a broader swathe of how complaint generators pursue takedowns. 
%Second, Lumen itself is biased towards a certain body of complaints (\ie DMCA notices). Thus, it is impossible to quantify how representative our analysis is of the wider complaint ecosystem. 
Our analysis has also revealed traits of a cat-and-mouse game, with complainants bulk sending notices, and websites replicating themselves across multiple domains and TLDs. 
Exploring the temporal attributes of this game will no doubt reveal a number of yet unseen behaviours. 
Last, we also wish to explore if search engines cease to index URLs that are complained about. Quantifying this forms a key strand of our future work.

% We also note that there are a number of limitations within our study, that provide the basis for several interesting lines of future work. 
% First, we emphasise that the Lumen dataset does not contain complaints made \emph{to} websites but, rather, \emph{about} websites, \eg requesting that Google remove a certain URL from their search results. Hence, we cannot definitively state that a website owner reacts to or ignores notices. 
% A future line of work is exploring exactly the types of notices that are issued to websites, and to capture a broader swathe of how complaint generators pursue takedowns. 
% Second, Lumen itself is biased towards a certain body of complaints (\ie DMCA notices). Thus, it is impossible to quantify how representative our analysis is of the wider complaint ecosystem. 
% An important line of future research is to compliment this with alternative measures of complaints activity. 
% Finally, we intent to take a more focused look at the actions and evasion tactics employed by website operators. We anticipate that a number of strategies are employed to evade both detection and takedown. Quantifying this forms a key strand of our future work.

%%%%%%%%%%%%%%%%%%%%%%%%%%% ACTUAL TEXT: END %%%%%%%%%%%%%%%%%%%%%%%%%%%%
\bibliographystyle{ACM-Reference-Format}
\balance
\bibliography{main}

\end{document}